\documentclass{aa}

\usepackage{graphicx}
\usepackage{txfonts}
\usepackage{hyperref}
\usepackage{multirow}
\usepackage{lscape}
\usepackage{amsmath}

\usepackage{xcolor}

\newcommand{\corr}{\mathrm{cor}}

\begin{document} 

 \title{Redshifting galaxies from DESI to JWST CEERS: Correction of biases and uncertainties in quantifying morphology}

   \author{Si-Yue Yu\inst{\ref{mpifr}}\thanks{Humboldt Postdoctoral Fellow},  
                        Cheng Cheng\inst{\ref{NAOC}},
           Yue Pan\inst{\ref{UChicago}},
           Fengwu Sun\inst{\ref{SO}},
           \and  
           Yang A. Li\inst{\ref{PKU}}
          }

   \institute{ 
   Max-Planck-Institut für Radioastronomie, Auf dem Hügel 69, 53121 Bonn, Germany\\ \email{syu@mpifr-bonn.mpg.de}\label{mpifr} 
   \and Chinese Academy of Sciences South America Center for Astronomy, National Astronomical Observatories, CAS, Beijing 100101, People's Republic of China \label{NAOC}
   \and Department of Astronomy \& Astrophysics, University of Chicago, 5640 South Ellis Avenue, Chicago, IL 60637, USA \label{UChicago}
   \and Steward Observatory, University of Arizona, 933 N. Cherry Avenue, Tucson, AZ 85721, USA \label{SO}
   \and Department of Astronomy, School of Physics, Peking University, Beijing 100871, China \label{PKU}
            }

\abstract{
Observations of high-redshift galaxies with unprecedented detail have now been rendered possible with the {\it James Webb} Space Telescope (JWST). However, accurately quantifying their morphology remains uncertain due to potential biases and uncertainties. To address this issue, we used a sample of 1816 nearby DESI galaxies, with a stellar mass range of $10^{9.75 \text{\textendash}11.25}$\,$M_{\odot}$, to compute artificial images of galaxies of the same mass located at $0.75\leq z\leq 3$ and observed at rest-frame optical wavelength in the Cosmic Evolution Early Release Science (CEERS) survey. We analyzed the effects of cosmological redshift on the measurements of Petrosian radius ($R_p$), half-light radius ($R_{50}$), asymmetry ($A$), concentration ($C$), axis ratio ($q$), and Sérsic index ($n$). Our results show that $R_p$ and $R_{50}$, calculated using non-parametric methods, are slightly overestimated due to PSF smoothing, while $R_{50}$, $q$, and $n$ obtained through fitting a Sérsic model does not exhibit significant biases. By incorporating a more accurate noise effect removal procedure, we improve the computation of $A$ over existing methods, which often overestimate, underestimate, or lead to significant scatter of noise contributions. Due to PSF asymmetry, there is a minor overestimation of $A$ for intrinsically symmetric galaxies. However, for intrinsically asymmetric galaxies, PSF smoothing dominates and results in an underestimation of $A$, an effect that becomes more significant with higher intrinsic $A$ or at lower resolutions. Moreover, PSF smoothing also leads to an underestimation of $C$, which is notably more pronounced in galaxies with higher intrinsic $C$ or at lower resolutions. We developed functions based on resolution level, defined as $R_p/$FWHM, for correcting these biases and the associated statistical uncertainties. Applying these corrections, we measured the bias-corrected morphology for the simulated CEERS images and we find that the derived quantities are in good agreement with their intrinsic values -- except for $A$, which is robust only for angularly large galaxies where $R_p/{\rm FWHM}\geq 5$. Our correction functions can be applied to other surveys, offering valuable tools for future studies.
}

\keywords{  galaxies: structure --
						galaxies: fundamental parameters --
                        galaxies: evolution --
                        galaxies: high-redshift
                        }
\maketitle

\section{Introduction}

Galaxy morphology has been traditionally described in a qualitative way using the {\it Hubble} sequence \citep{Hubble1926}, which is widely recognized as a fundamental aspect in the study of galaxy formation and evolution. With its unprecedented sensitivity and resolution in the infrared, the {\it James Webb} Space Telescope (JWST) is making significant advances in our understanding of the origin of the Hubble sequence. Previous studies using the {\it Hubble} Space Telescope (HST) suggested that the majority of galaxies at $z>2$ are peculiar \citep[e.g.,][]{Conselice2008, Mortlock2013}. However, early JWST studies reveal a large fraction of regular disk galaxies at high redshift \citep{Ferreira2022a, Ferreira2022b, Kartaltepe2023, Nelson2022, Robertson2023, Jacobs2023, Cheng2023, Cheng2022b}. These high-redshift disk galaxies can have spiral arms and bars similar to those in Local Universe \citep{Wu2022, Chen2022, Fudamoto2022, Guo2023}. Galaxies with established disk and spheroidal morphologies span the full redshift range \citep{Kartaltepe2023} and the Hubble Sequence was already in place as early as $z\approx 6$ \citep{Ferreira2022b}.

Quantifying morphology is crucial for exploring galaxy evolution and it can be achieved using both non-parametric and parametric methods \citep[e.g.,][]{Elmegreen1985, Conselice2003, Lotz2004, Elmegreen2007, Martin2014, YuHo2018, Yu2019, YuHo2020, Yu2021, Smith2022, Yu2022a, Yu2022b}.  The Cosmic Evolution Early Release Science (CEERS) survey (PI: Finkelstein, ID=1345, \citealt{Finkelstein2022a}; \citealt{Bagley2023}) is an early release science program in Cycle 1 that observes the Extended Growth Strip field (EGS) of the Cosmic Assembly Near-IR Deep Extragalactic Legacy Survey \citep[CANDELS;][]{Grogin2011, Koekemoer2011}. The CEERS will observe ten pointings with the Near-Infrared Camera \citep[NIRCam;][]{Rieke2023}, covering a total of 100 sq. arcmin. By quantifying morphology of galaxies from the first four pointings, \cite{Ferreira2022b} and \cite{Kartaltepe2023} show that spheroids exhibit a higher average Sérsic index, smaller size, and rounder shape compared to disks and peculiars, although selection effects may exist. Additionally, the average Sérsic index decreases with increasing redshift \citep{Ferreira2022b}. Despite the slightly higher average concentration in spheroids and slightly higher average asymmetry in peculiars, the concentration-asymmetry diagram does not provide a clear separation of galaxies according to their morphological types \citep{Ferreira2022b, Kartaltepe2023}.

However, these early results may not accurately reflect the intrinsic galaxy morphology as the physical resolution and signal-to-noise ratio (S$/$N) of the JWST images of high-redshift galaxies are limited. Furthermore, comparing galaxy morphologies at different redshifts and/or observed by different instruments is challenging as changes in resolution, noise level, and rest-frame wavelength can alter the morphology and introduce biases and uncertainties in the quantification.

A commonly used strategy for understanding measurement biases and uncertainties caused by image degradation is to use high-quality images of low-redshift galaxies to generate simulated images of high-redshift galaxies and, subsequently, to compare the measurements before and after the image simulation. The pioneering work in this area was done by \cite{Giavalisco1996}. This strategy has been used to study spiral structure \citep[e.g.,][]{Block2001, vandenBergh2002, Yu2018}, bar structure \cite[e.g.,][]{vandenBergh2002, Sheth2008}, concentration-asymmetry-smoothness statistic \citep[e.g.,][]{Abraham1996, Conselice2003, Lee2013, Yeom2017, Whitney2021}, Gini-$M_{20}$ statistic \citep[e.g.,][]{Lee2013, Petty2014}, and Sérsic index and size \citep[e.g.,][]{Barden2008, Paulino-Afonso2017} of high-redshift galaxies observed by HST. Despite the widespread use of this technique, their image simulation procedures did not take into account the intrinsic galaxy size evolution and cannot be directly applied to galaxies with high redshifts. It has been found that there is a strong redshift evolution in galaxy size \citep[e.g.,][]{Bouwens2004, Daddi2005, Trujillo2007, Buitrago2008, Oesch2010, Mosleh2012, vanderWel2014, Whitney2019}. By using data from CANDELS, \cite{vanderWel2014} show that galaxies of a given stellar mass are on average smaller at higher redshifts, with fast evolution for early-type galaxies and moderate evolution for late-type galaxies.

Building on prior studies and by taking into account all relevant factors of image simulation, our goal is to use high-resolution and high-S$/$N nearby galaxy images to generate artificial images of galaxies located at redshift $0.75\leq z\leq3$ and observed at optical rest-frame wavelength in JWST CEERS. We then go on to investigate biases and uncertainties in the quantification of galaxy morphology, focusing on six key morphological quantities: Petrosian radius, half-light radius, asymmetry, concentration, Sérsic index, and axis ratio, which are commonly used to describe the typical morphology of a galaxy. We aim to derive corrections to these biases and uncertainties to improve the accuracy and robustness of future galaxy morphology studies.

This paper is organized as follows. Section~\ref{data} outlines our sample selection of nearby galaxies and the data reduction process. Section~\ref{Art} provides a detailed description of the image simulation methodology. Section~\ref{rob} discusses the biases and uncertainties, and derive correction functions. Section~\ref{validation} validates the effectiveness of the correction functions. Finally, a summary of the main findings is presented in Sect.~\ref{conclusions}. Throughout this work, we use AB magnitudes and assume the following cosmological parameters: $(\Omega_{\rm M}, \Omega_{\Lambda}, h)=(0.27, 0.73, 0.70)$.

\section{Observational material}\label{data}

\begin{figure*}
        \centering
        \includegraphics[width=0.95\textwidth]{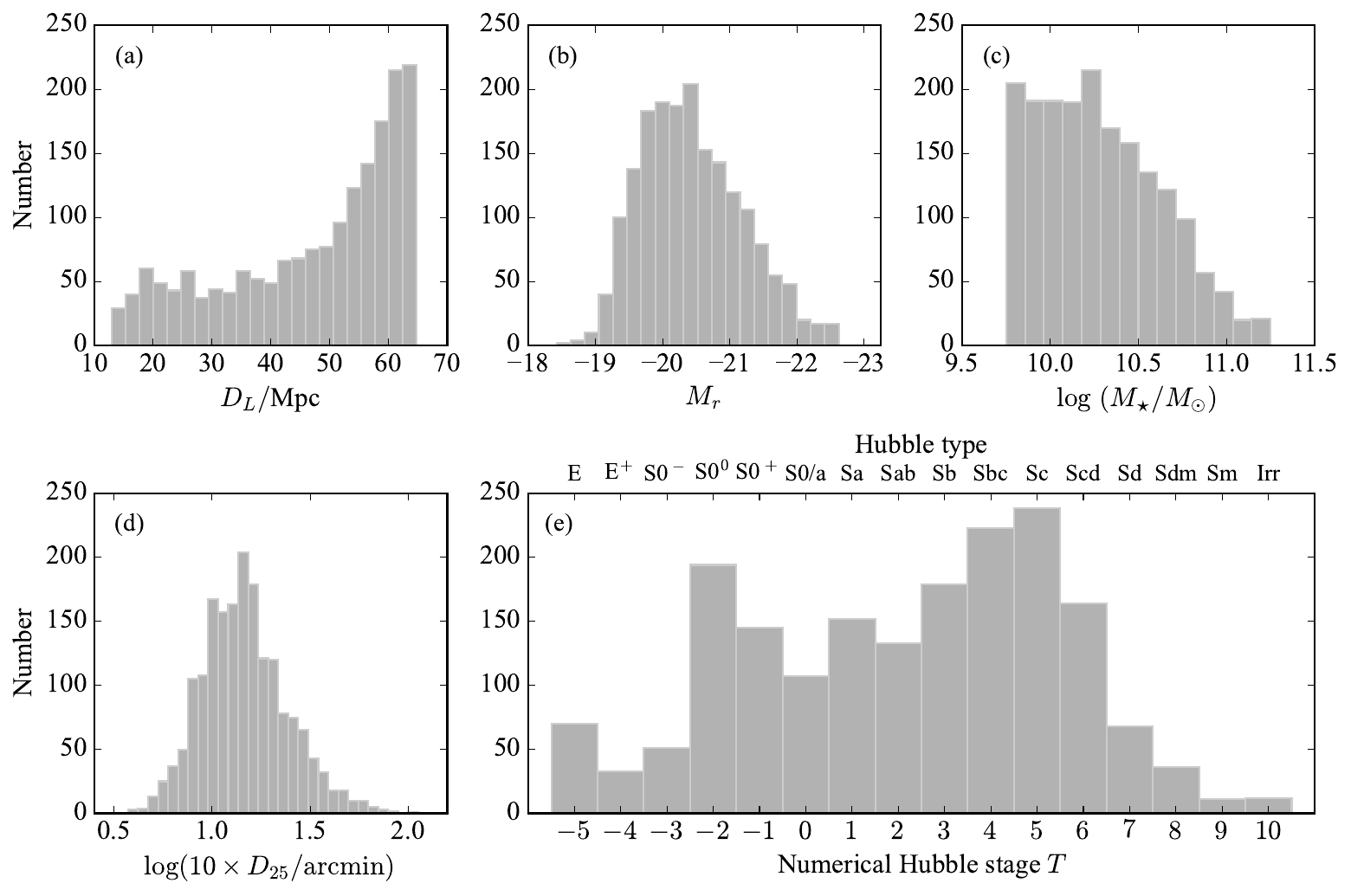}
        \caption{Basic properties of our sample. Distribution of (a) luminosity distances, (b) absolute {\it r}-band magnitude, corrected for Galactic extinction, (c) stellar mass, (d) {\it B}-band isophotal diameter at $\mu_B=25$\,mag\,arcsec$^{-2}$, and (e) Hubble types.
        }
        \label{basic}
\end{figure*}

\subsection{Sample of nearby galaxies}

We restricted the redshift range of our image simulations to $z=0.75$, 1.0, 1.25, ..., and 3.0. We refrain from simulating images of galaxies at $z>3$, as the of evolution galaxy optical size at these redshifts is not yet well constrained. For each target redshift, we chose the filter that observes the rest-frame optical wavelength ($\lambda=5000$\textendash$7000\,\AA$). Specifically, we used the F115W filter for $z=0.75$ and 1, the F150W filter for $z=1.25$, 1.5, and 1.75, and the F200W filter for $z=2.0$ to 3.0. To generate artificially redshifted galaxy images, nearby ($z\approx0$) galaxy images observed in a similar rest-frame wavelength range are required. This range is covered by the {\it g}, {\it r}, and {\it z} bands (with effective wavelengths: 4796\,\AA, 6382\,\AA, and 9108\,\AA) provided by the Dark Energy Spectroscopic Instrument (DESI) Legacy Imaging Surveys\footnote{http://legacysurvey.org/} \citep[][reference therein]{Dey2019}.

The DESI Legacy Imaging Surveys are comprised of three public projects: the Beijing-Arizona Sky Survey \citep[BASS;][]{zou2017}, the Mayall {\it z}-band Legacy Survey \citep[MzLS;][]{Blum2016}, and the Dark Energy Camera Legacy Survey \citep[DECaLS;][]{Blum2016}. We focus on the DECaLS {\it g}-, {\it r}, and {\it z}-band images, while the MzLS {\it z}-band nearby galaxy images are found to be significantly affected by pattern noise and are therefore excluded from our analysis along with their corresponding BASS {\it g}- and {\it r}-band images. We defined our sample of nearby galaxies using the Siena Galaxy Atlas (SGA)\footnote{https://www.legacysurvey.org/sga/sga2020/}, which is constructed based on the DESI and includes 383,620 galaxies. The SGA does not include small objects with $D_{25}<20$\,arcsec, where $D_{25}$ is the {\it B}-band isophotal diameter at $\mu_B=25$\,mag\,arcsec$^{-2}$, as a high fraction of them are spurious sources.

We selected galaxies with available Hubble type classification and best-estimated luminosity distance ($D_L$) from HyperLeda extragalactic database\footnote{http://leda.univ-lyon1.fr/} \citep[][reference therein]{Makarov2014}. The best-estimated distance is weighted average between distance derived from spectroscopic redshift and published redshift-independent distances, and it provide an homogeneous distance estimate over the whole redshift range \citep{Makarov2014}. We select galaxies with $12.88\leq D_L \leq 65.01$\,Mpc, corresponding to cosmological $z=$ 0.003\textendash0.015. We exclude galaxies in the Galactic plane ($-20\degr \leq b\leq 20\degr$) to avoid any possible severe photometric problems caused by crowd foreground stars. We excluded galaxies that are covered by masks of nearby bright sources provided by the SGA, as the emission from these galaxies is severely contaminated by the bright sources. This process will remove merging systems where the centers of two galaxies are close but have not yet merged into one. The SGA catalog also includes mid-infrared photometry from the {\it Wide-field Infrared Survey Explorer} \citep[WISE;][]{Wright2010}. We use the DECaLS {\it g}-, {\it r}-, and {\it z}-band flux and WISE W1 flux to estimate stellar mass ($M_\star$) through SED fitting using {\tt CIGALE}\footnote{https://cigale.lam.fr/; we use 2022 version} \citep{Boquien2019}, assuming the Chabrier stellar initial mass function \citep{Chabrier2003}, double exponential star formation history, simple stellar population of \cite{bc03}, and attenuation law of \cite{Calzetti2000}. We selected galaxies with $M_\star$ of $10^{9.75 \text{\textendash}11.25}$\,$M_{\odot}$. The mass cut was applied because we use the galaxy size evolution derived by \cite{vanderWel2014}, which is only available in this mass range for both early- and late-type galaxies. Our final sample of nearby galaxies consists of 1816 galaxies.

Figure~\ref{basic} summarizes some of the basic parameters of the sample. Most of the galaxies are nearby (median $D_L=52.2$\,Mpc; Fig.~\ref{basic}(a)), luminous (median $M_r=-20.4$\,mag, corrected for Galactic extinction; Fig.~\ref{basic}(b)), massive (median $M_{\star}=10^{10.2}\,M_{\odot}$; Fig.~\ref{basic}(c)), and angularly large (median $D_{25}=1.4$\,arcmin; Fig.~\ref{basic}(d)).  The sample spans the full range of Hubble types in the nearby Universe (Fig.~\ref{basic}(e)), comprising 600 (33\%) early-type galaxies (ellipticals, S0, or S0/a), and 1216 (67\%) late-type galaxies (Sa\textendash Irr). We acquired {\it g}-, {\it r}-, and {\it z}-band mosaic images and point spread functions (PSFs) from SGA. The Galactic extinction is corrected using the map of dust reddening \citep{SFD98, SF2011}. We use the galaxy center, ellipticity, and position angle provided by the SGA catalog when removing foreground stars (as described in Sect.~\ref{remove}). We set $R_{25}=D_{25}/2$. We measured the sky background and noise using {\tt AutoProf}\footnote{https://autoprof.readthedocs.io/} \citep{Stone2021}. The sky background is then subtracted from the image.

\subsection{Removal of foreground stars}\label{remove}
Contamination from sources other than the target galaxy, such as foreground stars and projected close galaxies, should be removed or minimized prior to image simulation. This process was not done in some previous studies on simulating high-redshift galaxy images observed by HST (e.g., \citealt{Barden2008}; \citealt{Paulino-Afonso2017}, but also see \citealt{Yu2018}), rendering their results less robust. We first mask out the contamination. For each galaxy image, the SGA provides a catalog of sources  that are identified using {\it The Tractor}\footnote{https://github.com/dstndstn/tractor/}, a forward-modeling approach to performing source extraction and model fitting to the sources. We masked out each identified star, using a mask size determined by the star-centered radius at which the {\it r}-band flux start to fluctuate due to galaxy structure or background noise. We then masked out each identified projected nearby galaxy using an ellipse with semi-major axis of 1.5\,$R_{25}$. Finally, we performed a visual inspection and manually masked out any residual stars or small background galaxies that were missed in the above process. We adopted the same mask for {\it g}-, {\it r}-, and {\it z}-band images.

We removed contamination through the following steps. The masked regions outside the galaxy ($R>1.5\,R_{25}$) are set to zero. For small masked regions (area $<$\,50\,arcsec$^2$) inside the galaxy ($R\leq 1.5\,R_{25}$), we estimated the intrinsic galaxy light affected by foreground stars and/or projected close galaxies using interpolation. For each masked region, we cut out a square region containing twice as many unmasked pixels as masked pixels. The interpolation was then performed by approximating the values of the unmasked pixels by a polynomial function. We used the interpolated values to fill in the small masked regions.

For large masked regions (area $\geq$\,50\,arcsec$^2$) inside the galaxy, interpolation may fail to reproduce the intrinsic galaxy light, as the complex galaxy structures may cause overfitting and lead to catastrophic results. Instead, we replaced the masked region with values from their 180$\degr$ rotational symmetric pixels if the symmetric portion did not have a large masked region. The rotational symmetric images were originally used to highlight spiral structure \citep{Elmegreen1992, Elmegreen2011}. We caution that this approach may not restore the flux in prominent three-armed structures, which are 120$\degr$ rotational symmetric, but this effect is small since the fraction of three-armed structures is small \citep{Elmegreen1987}. In cases where the 180$\degr$-rotational symmetric portion also has a large masked region or the galaxy is highly inclined ($i>70\degr$), we filled in the masked region with values from their mirror-symmetric pixels reflected over the galaxy major axis. In a few instances where neither of the above criteria are satisfied, we reverted to the interpolation method and used a low-order polynomial function to perform fitting and interpolation. Finally, we added Poisson noise to the cleaned regions to simulate real observations.

The above process makes use of galaxy symmetry, but it does not significantly affect the calculation of galaxy asymmetry described in Sect.~\ref{rob}, as the masked regions are small relative to the galaxy size. We have to sacrificed a small degree of precision in computing the galaxy asymmetry to generate cleaned images for image simulation.

\begin{figure}
        \centering
        \includegraphics[width=0.49\textwidth]{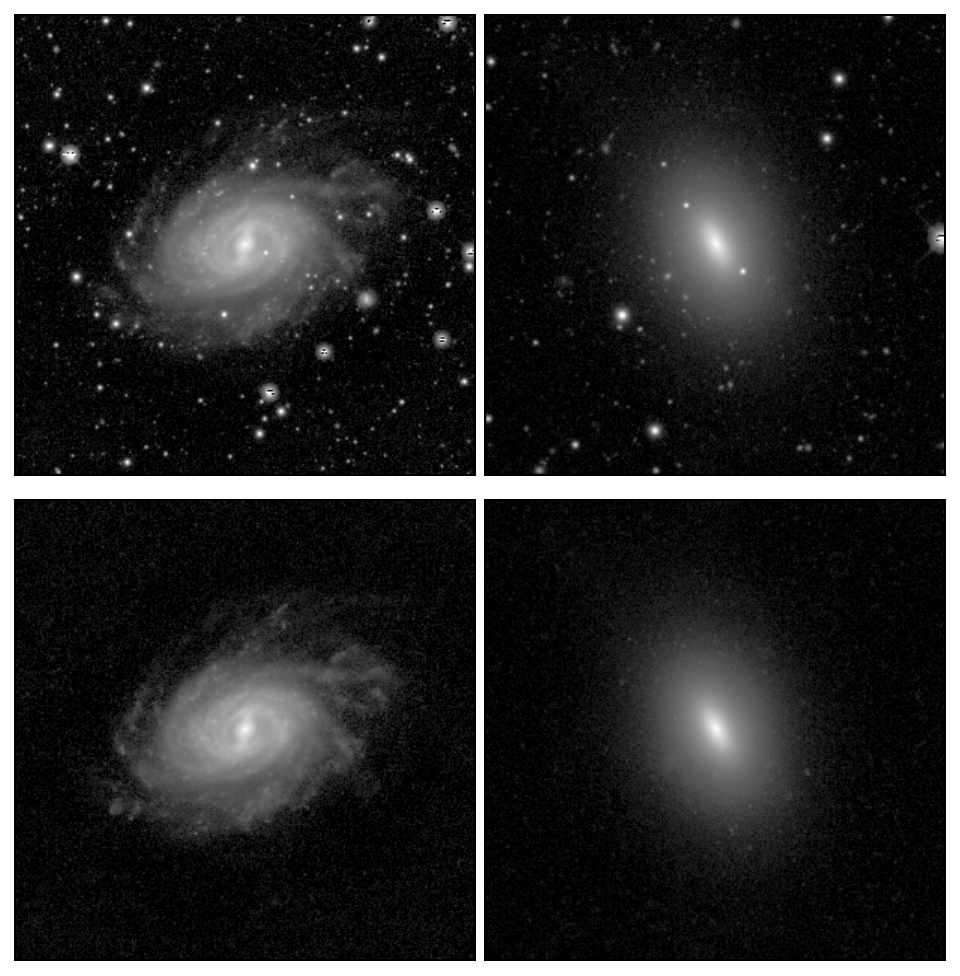}
        \caption{Demonstration of our cleaning process. Left and right columns showcase the {\it r}-band observation of two galaxies, ESO~121-026 (a Sbc galaxy) and ESO~251-004 (an elliptical galaxy). Top and bottom rows  display the image before and after undergoing star removal, respectively. The vast majority of emission from foreground stars or background galaxies in the vicinity of the galaxies was successfully removed. 
        }
        \label{clean}
\end{figure}

The removal of foreground stars is done for each galaxy to generate their star-cleaned {\it g}-, {\it r}-, and {\it z}-band images. The effectiveness of the cleaning is illustrated in Fig.~\ref{clean}, showing the {\it r}-band images of two galaxies, ESO~121-026 (a Sbc galaxy) and ESO~251-004 (an elliptical galaxy), before and after undergoing the removal process. Our process successfully eliminates the vast majority of contamination surrounding the galaxies, revealing clearer and more accurate images of the galaxy structures. The median PSF FWHM of DECaLS images is $\sim$\,1.3, 1.2, and 1.1 arcsec in the {\it g}, {\it r}, and {\it z} bands, respectively \citep{Dey2019}. To facilitate the pixel-by-pixel {\it K} correction described in Sect.~\ref{RP}, we matched the {\it g}-, {\it r}-, and {\it z}-band images to a common PSF for each galaxy. We did not use Fourier transformation to find a convolution kernel, because this would introduce significant high frequency noise, as the PSF FWHMs at different bands are very close. Instead, the matching was done by searching for a kernel of Moffat function, which is convolved with the PSF of smaller FWHM to get a broadened PSF that has almost the same best-fitted Moffat function with the PSF of larger FWHM. The star-cleaned {\it g}-, {\it r}-, and {\it z}-band images with a common PSF were used to compute artificial high-redshift galaxy images observed in JWST CEERS.

\section{Artificially redshifting galaxies} \label{Art}

The nearby DESI galaxy images are of fairly high quality, which allows for accurate determinations of morphological measurements. Compared to DESI images, JWST CEERS images of high-redshift galaxies have lower physical resolution and lower S$/$N, as the galaxies become less resolved and fainter at higher redshifts. The limited data quality may bias measurements and make them more uncertain. To understand the how well we can quantify galaxy morphology using the JWST NIRCam images, we simulated nearby DESI galaxies with respect to how they would appear at various high redshifts, as observed in JWST CEERS. Section~\ref{formulae} presents the derivation of the formulae used in the redshifting procedure. Subsequently, Sections~\ref{SE} and~\ref{LE} respectively discuss the evolution of galaxy size and luminosity. The redshifting procedure is outlined in Sect.~\ref{RP}.

\subsection{Formulae}\label{formulae}
We begin by assuming the existence of an extended source located at redshift, $z_{\rm local}$. Its luminosity distance and flux density we observed are denoted as $D_L(z_{\rm local})$ and $f_{\rm local}$, respectively. We adopted erg\,s$^{-1}$\,cm$^{-2}$\,Hz$^{-1}$ as the unit of $f$ because we use AB magnitude. The energy emitted by the source in 1\,second at a narrow range of wavelength from $\lambda-\Delta\lambda/2$ to $\lambda+\Delta\lambda/2$ is given by:
\begin{equation} \label{flocal}
  E_{\rm local} = 4\pi\cdot \frac{c}{\lambda^2}\cdot f_{\rm local}\cdot D_L^{2}(z_{\rm local})\cdot \Delta\lambda,
\end{equation}
\noindent
where $c$ is the light speed. The source is manually moved to a higher redshift of $z_{\rm high}$ with the luminosity distance of $D_L(z_{\rm high})$. The observed flux density is denoted as $f_{\rm high}$. The observed wavelength becomes $\lambda^\prime = \lambda\cdot (1+z_{\rm high})/(1+z_{\rm local})$ and the wavelength width becomes $\Delta\lambda^\prime = \Delta\lambda \cdot (1+z_{\rm high})/(1+z_{\rm local})$. The energy emitted by the source in 1\,second at a narrow range of wavelength from $\lambda^\prime-\Delta\lambda^\prime/2$ to $\lambda^\prime+\Delta\lambda^\prime/2$ is:
\begin{equation}\label{fhig}
  E_{\rm high} = 4\pi\cdot \frac{c}{\lambda^{\prime 2}}\cdot f_{\rm high}\cdot D_L^{2}(z_{\rm high})\cdot \Delta\lambda^\prime.
\end{equation}
\noindent
We assume that the source becomes intrinsically brighter at $z_{\rm high}$, with the energy emitted given by:

\begin{equation}\label{engergy}
  E_{\rm high} = \left(\frac{1+z_{\rm high}}{1+z_{\rm local}}\right)^{\alpha} \cdot E_{\rm local}.
\end{equation}

\noindent
Combining Eq.~(\ref{flocal}), (\ref{fhig}), and (\ref{engergy}), we obtain:
\begin{equation}\label{fscale}
  f_{\rm high} = \frac{D^2_L(z_{\rm local})}{D^2_L(z_{\rm high})}\cdot \frac{1+z_{\rm high}}{1+z_{\rm local}} \cdot \left(\frac{1+z_{\rm high}}{1+z_{\rm local}}\right)^{\alpha} \cdot f_{\rm local}.
\end{equation}
\noindent
The first term on the right-hand side of the equation is caused by the distance effect and the second term is caused by the cosmological compression of the frequency (or, equivalently, the cosmological dilation of the wavelength). The second term occurs as we consider monochromatic luminosity, while it would disappear if we considered bolometric luminosity. The third term is caused by the assumed luminosity evolution. Equation~(\ref{fscale}) is used to rescale the flux in the redshifting procedure (Sect.~\ref{RP}). Hence, we can relate the absolute magnitude of the source at local and high redshift using:
\begin{equation}\label{Mevo}
  M_{\rm high} =  - 2.5\log \left( \frac{1+z_{\rm high}}{1+z_{\rm local}} \right)^{\alpha} + M_{\rm local}.
\end{equation}

We assume the source has a physical radius of $R_{\rm local}$ at $z_{\rm local}$ and $R_{\rm high}$ at $z_{\rm high}$. The solid angle it spans at $z_{\rm local}$ and $z_{\rm high}$ is: 
\begin{equation}\label{gloc}
  \Omega_{\rm local}=\frac{\pi R_{\rm local}^2}{D^2_{\rm ang}(z_{\rm local})}
,\end{equation}

\noindent
and

\begin{equation}\label{ghig}
  \Omega_{\rm high}=\frac{\pi R_{\rm high}^2}{D^2_{\rm ang}(z_{\rm high})}, 
\end{equation}

\noindent
respectively. $D_{\rm ang}(z)$ is the angular-diameter distance at redshift $z$. We assume the physical size becomes smaller at higher redshift: 
\begin{equation}\label{Revo}
  R_{\rm high} = \left(\frac{1+z_{\rm high}}{1+z_{\rm local}}\right)^{\beta}\cdot R_{\rm local}.
\end{equation}

\noindent
Combining Eq.~(\ref{gloc}), (\ref{ghig}), and (\ref{Revo}), we obtain:
\begin{equation} \label{SizeE}
  \Omega_{\rm high} = \frac{D^2_{\rm ang}(z_{\rm local})  } {D^2_{\rm ang}(z_{\rm high})} \cdot  \left( \frac{1+z_{\rm high}}{1+z_{\rm local}} \right)^{2\,\beta} \cdot \Omega_{\rm local}
.\end{equation}
\noindent
We denote the pixel scale as $p$ and number of pixels occupied by the source as $N$. We therefore have:
\begin{equation} \label{NE}
  \frac{N_{\rm high}}{N_{\rm local}} = \frac{D^2_{\rm ang}(z_{\rm local})  } {D^2_{\rm ang}(z_{\rm high})} \cdot  \left( \frac{1+z_{\rm high}}{1+z_{\rm local}} \right)^{2\,\beta} \cdot \frac{p^2_{\rm local}}{p^2_{\rm high}},
\end{equation}

\noindent
which is the binning factor we use to downscale image size in the redshifting procedure (Sect.~\ref{RP}). The flux density per unit solid angle can be written as:
\begin{equation}
  \frac{f_{\rm high}}{\Omega_{\rm high}} = \left( \frac{1+z_{\rm high}}{1+z_{\rm local}} \right)^{-3} \cdot \left(\frac{1+z_{\rm high}}{1+z_{\rm local}}\right)^{\alpha} \cdot \left( \frac{1+z_{\rm high}}{1+z_{\rm local}} \right)^{-2\,\beta} \cdot
  \frac{f_{\rm local}}{\Omega_{\rm local}}
.\end{equation}

\noindent
Denoting the surface brightness as $\mu$, we have:
\begin{multline}
  \mu_{\rm high} = \mu_{\rm local} + 2.5\log \left( \frac{1+z_{\rm high}}{1+z_{\rm local}} \right)^3 
  -2.5 \log \left( \frac{1+z_{\rm high}}{1+z_{\rm local}} \right)^{\alpha} \\
  -2.5 \log \left( \frac{1+z_{\rm high}}{1+z_{\rm local}} \right)^{-2\,\beta}.
\end{multline}

\noindent
If we adopt $z_{\rm local}=0$ and $z_{\rm high}=z$, we have:
\begin{equation}\label{SBevo2}
  \mu_z  = \mu_0 + 2.5\log \left( 1+z \right)^3 - 2.5 \log \left( 1+z \right)^{\alpha-2\beta}.
\end{equation}

\noindent
The term of $2.5\log \left( 1+z \right)^3$ is the well-known cosmological dimming, while the term of $2.5 \log \left( 1+z \right)^{\alpha-2\beta}$ describes the evolution of surface brightness. In the redshifting procedure, we assumed that the average flux density observed in a specific filter is consistent with the flux density at the effective wavelength of the filter, so that we can ignore the difference between bandpass width of the DESI filter and JWST filter.

\subsection{Size evolution}\label{SE}

Galaxies at a fixed mass are physically smaller at higher redshift \citep[e.g.,][]{Bouwens2004, Daddi2005, Trujillo2007, Buitrago2008, Oesch2010, Mosleh2012, vanderWel2014, Whitney2019}. Using structural parameters derived from CANDELS imaging \citep{Grogin2011, Koekemoer2011}, \cite{vanderWel2014} found a significantly different rate ($\beta$) of average size evolution for early-type and late-type galaxies. The average effective radius of early-type galaxies, calculated over a range of stellar masses, evolves rapidly, following $R_{\rm eff}\propto(1+z)^{\beta=-1.48}$, while that of late-type galaxies evolves moderately, following $R_{\rm eff}\propto(1+z)^{\beta=-0.75}$. Therefore, when simulating high-redshift galaxies, it is important to properly account for intrinsic size evolution; otherwise, the simulated galaxies would be larger than the true galaxies of the same mass.

The evolution rate is dependent on stellar mass, with more massive galaxies showing a higher rate (more negative $\beta$), as reported in \cite{vanderWel2014}. The dependence of $\beta$ on mass is weak for late-type galaxies, but strong for early-type galaxies, and less massive early-type galaxies evolve at a similar rate to late-type galaxies of the same mass. We adopted the $\beta$ measured by \cite{vanderWel2014}, as given in their Table~2) We focused on galaxies with a stellar mass of $10^{9.75 \text{\textendash}11.25}$\,$M_{\odot}$, so that  $\beta$ is available for both late-type and early-type galaxies. To estimate $\beta$ for our nearby DESI galaxies, we fit two polynomials to the $\beta$ as a function of stellar mass for late-type and early-type galaxies, respectively. These  best-fit functions were used to compute $\beta$ using stellar mass and galaxy type as input. The estimated $\beta$ was used for generating artificially redshifted galaxy images (see Sect.~\ref{RP}).  Additionally, galaxies tend to be larger at bluer wavelength than at redder wavelength \citep[e.g.,][]{Kelvin2012}. We performed a pixel-by-pixel {\it K} correction in Sect.~\ref{RP} to correct for this effect.

The rate of size evolution may differ when using different definitions of galaxy size. As shown by \cite{Whitney2019}, the outer radius of galaxies evolves at a faster rate than the inner radius, suggesting inside-out growth. While we do not consider the effect of inside-out growth in this study, we aim to compare our simulated images with real observations to investigate this effect in the future.

\begin{figure*}
        \centering
        \includegraphics[width=0.9\textwidth]{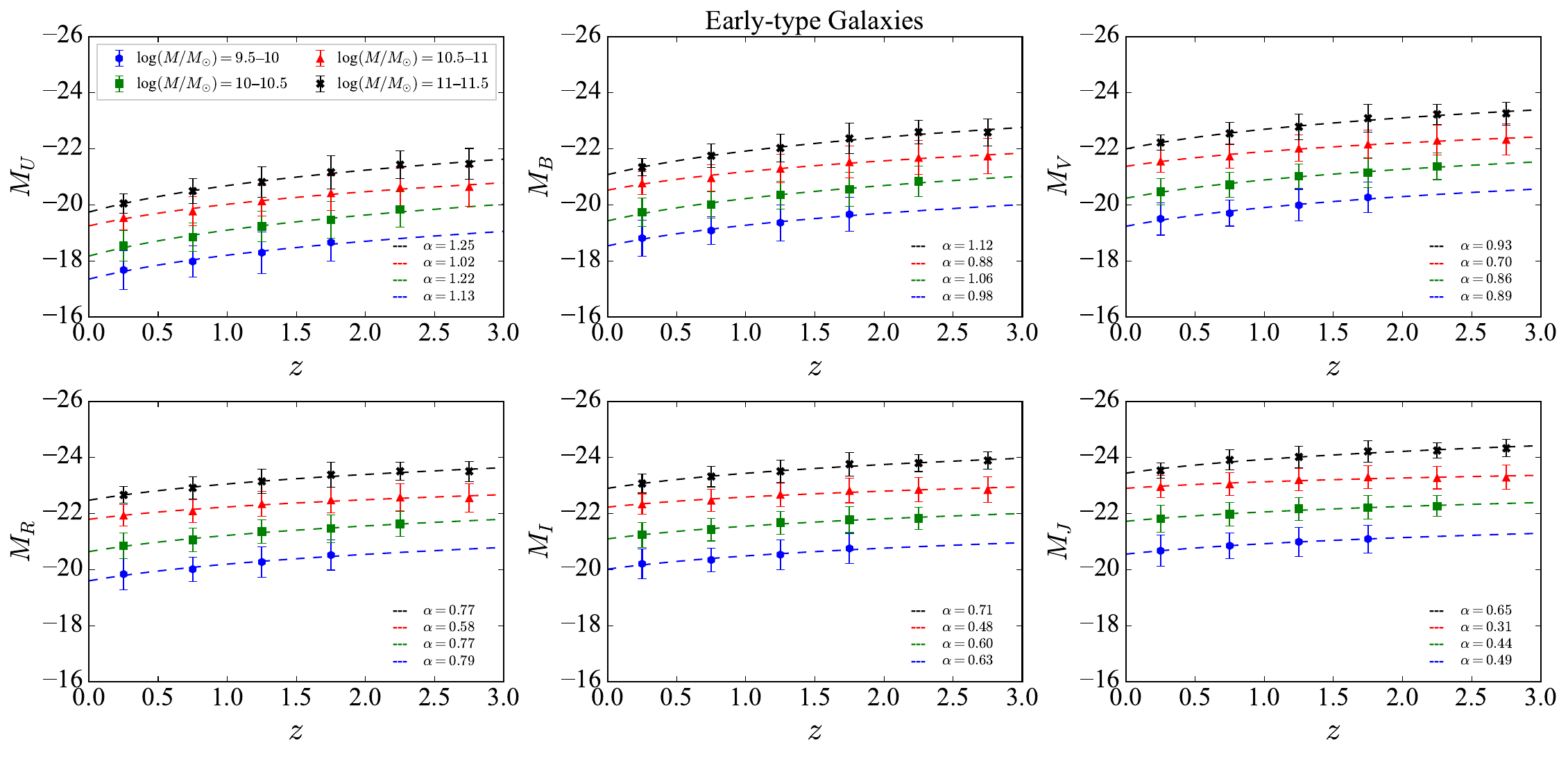}
        \caption{Redshift evolution of intrinsic absolute magnitude at rest-frame waveband of {\it U}, {\it B}, {\it V}, {\it R}, {\it I}, and {\it J} for early-type galaxies. The measured parameter $\alpha$ (Eq.~(\ref{Mfit})) describing the evolution rate for a given mass range and a given rest-frame waveband is shown in the bottom-right corner.
        }
        \label{M_evo_early}
\end{figure*}

\begin{figure*}
        \centering
        \includegraphics[width=0.9\textwidth]{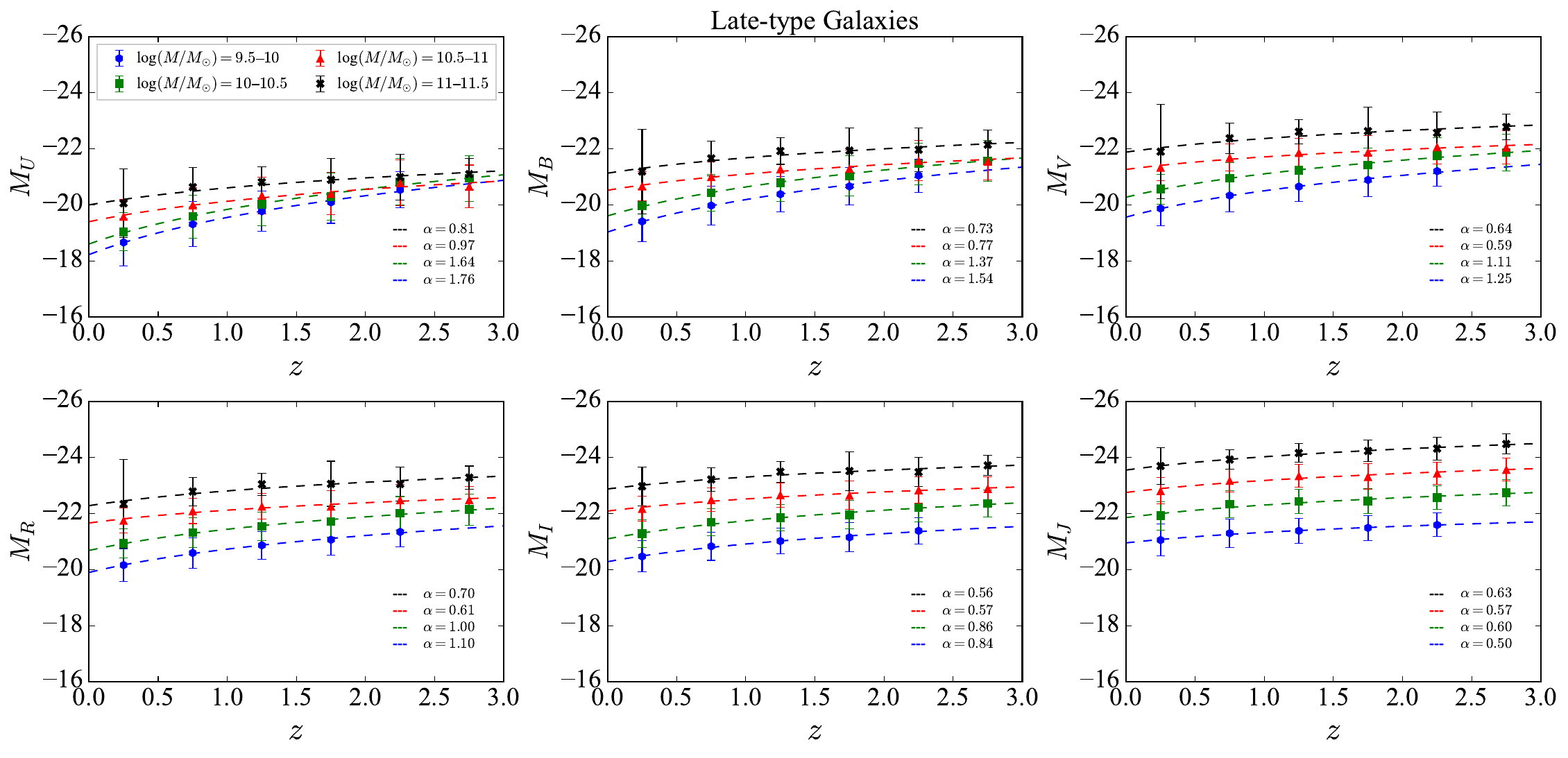}
        \caption{Same details as in Fig.~\ref{M_evo_early}, but for late-type galaxies.
        }
        \label{M_evo_late}
\end{figure*}

\subsection{Monochromatic luminosity evolution} \label{LE}
The intrinsic galaxy surface brightness has been found to brighten on average with increasing redshift \citep[e.g.,][]{Schade1995, Schade1996, Lilly1998, Roche1998, Labbe2003, Barden2005, Sobral2013, Whitney2020}. By definition, the evolution of surface brightness is partly attributed to intrinsic size evolution and partly attributed to monochromatic luminosity evolution. While the intrinsic size evolution is discussed in Sect.~\ref{SE}, in
this section we study  the monochromatic luminosity evolution by assuming a function form of $L_{\lambda}\propto (1+z)^{\alpha}$, where $L_{\lambda}$ is the monochromatic luminosity at rest-frame $\lambda$ filter with the effect of cosmological dimming corrected.

\begin{figure}
        \centering
        \includegraphics[width=0.45\textwidth]{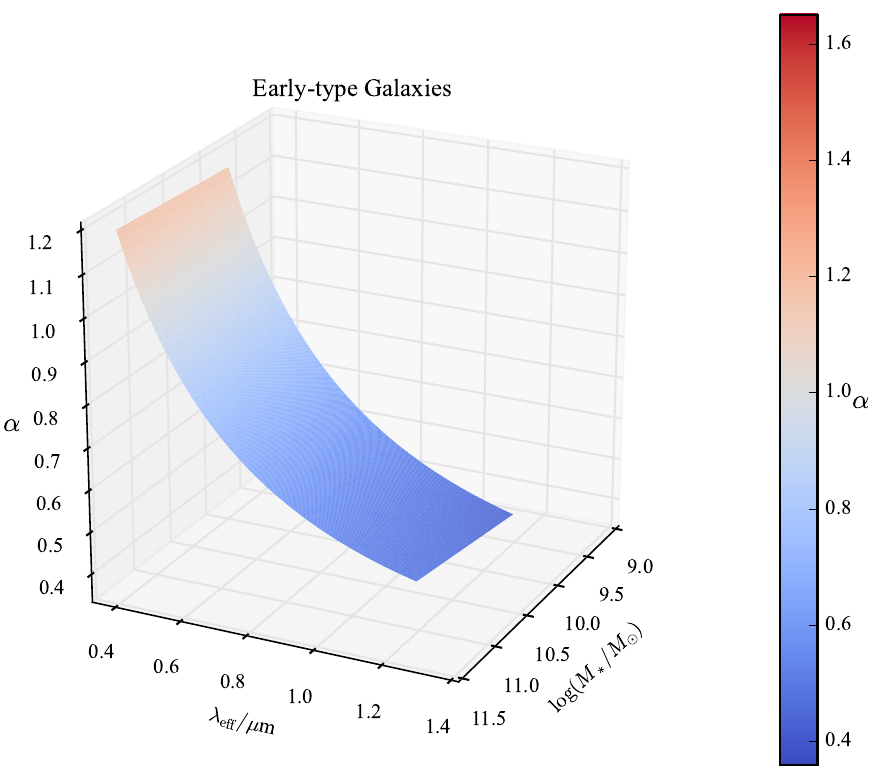}
        \caption{Best-fit $\alpha$ as a function of filter effective wavelength ($\lambda_{\rm eff}$) and stellar mass ($M_*$) for early-type galaxies. The color bar encodes $\alpha$. 
        }
        \label{a_early_3D}
\end{figure}

\begin{figure}
        \centering
        \includegraphics[width=0.45\textwidth]{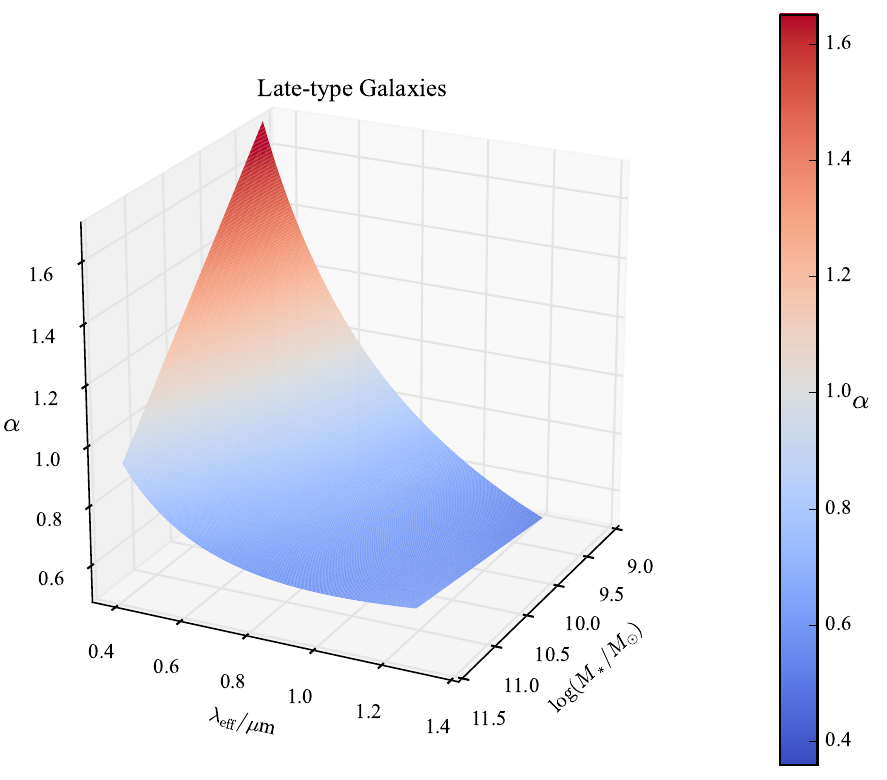}
        \caption{Same details as in Fig.~\ref{a_early_3D}, but for late-type galaxies.
        }
        \label{a_late_3D}
\end{figure}

To stay consistent with \cite{vanderWel2014}, we used the rest-frame flux, magnitude,  color index, and redshift from the 3D-HST catalog \citep{Brammer2012, Skelton2014}. Following the strategy in \cite{vanderWel2014}, we selected CANDELS galaxies with F160W apparent magnitude brighter than 25.5, a flag for good model fittings, and stellar mass above the mass of completeness limit at each redshift range. We classified the selected CANDELS galaxies into early-type and late-type galaxies using the demarcation lines proposed by \cite{Williams2009} in the diagram of $U-V$ versus $V-J$ color index. We separated the early-type and late-type galaxies to various bins of redshift ($\Delta z=0.5$) and stellar mass ($\Delta \log M_*=0.5$\,dex), as used in \cite{vanderWel2014}. Then we calculated the median rest-frame absolute magnitude $M_{\lambda}$, where $\lambda$ denotes {\it U}, {\it V}, {\it B}, {\it R}, {\it I,} or {\it J} filter. We fit the following function: 
\begin{equation} \label{Mfit}
  M_{\lambda}=M_{\lambda,0} - 2.5\,\log(1+z)^{\alpha}, 
\end{equation}
to the $M_{\lambda}$ as a function of $z$ for each bin of stellar mass viewed at each filter to determine $\alpha$. The results are shown in Figs.~\ref{M_evo_early} and \ref{M_evo_late}. We then fit two 2D polynomials to the measured $\alpha$ as a function of filter effective wavelength and stellar mass for early-type and late-type galaxies, respectively.  Figures~\ref{a_early_3D} and \ref{a_late_3D} show the best-fit 2D functions. The color bar encodes best-fit $\alpha$ for a given wavelength and mass. Both early-type and late-type galaxies get brighter on average at higher redshifts across all wavebands. The rate of evolution is more rapid in bluer wavebands than in redder ones, owing to more intense star formation in the past \citep[][]{Speagle2014, Scoville2023}. Interestingly, the low-mass late-type galaxies have particularly high $\alpha$ at blue wavelength. Our results are consistent with the evolution of the characteristic magnitude of luminosity function, where the characteristic magnitude becomes brighter at higher redshifts and evolves faster in the UV than in the {\it V} band. \citep[e.g.,][]{Arnouts2005, Marchesini2012}. We used the rest-frame wavelength, galaxy type, and stellar mass as inputs for estimate $\alpha$ for our nearby DESI galaxies by using the best-fit polynomials.

\subsection{Redshifting procedure} \label{RP}

Our redshifting procedure is based on the method of \cite{Giavalisco1996}, which we updated by incorporating the {\it K} correction, galaxy size evolution, and  luminosity evolution. This approach ensures that the artificially redshifted galaxy images closely match the size and brightness of true galaxies viewed at high redshifts. We compute artificial images of galaxies at $z=0.75$, 1.0, 1.25, 1.5, 1.75, 2.0, 2.25, 2.5, 2.75, and 3.0, using the F115W filter for $z=0.75$ and 1, the F150W filter for $z=1.25$, 1.5, and 1.75, as well as the F200W filter for $z=2.0$ to 3.0. To determine the typical noise level for the simulated images, we use the science data, error map, and source mask from the CEERS Data Release Version 0.5 provided by \cite{Bagley2023}.  In order to simulate background noise at each filter, we cut out a patch with a few sources from the CEERS science data, and then replaced these sources with background regions next to the sources to get a clean fragment of a real background image. We selected 80 galaxies, consisting of the top 20 brightest galaxies from each of the four pointings, to calculate a median ratio of galaxy flux variance to galaxy flux,  used to estimate the noise from galaxy flux. This quantity depends on filter, exposure time, sky brightness, and system throughput, and varies only slightly from location to location in CEERS. We adopted a pixel scale of 0.03\,arcsec/pixel, same as in \cite{Bagley2023}. We generated a two-time oversampling PSFs at the F115W, F150W, and F200W filters using {\tt WebbPSF} \citep{Perrin2014}. The two-time oversampling PSFs were used because the F115W and F150W PSFs on the pixel scale of 0.03\,arcsec/pixel were undersampled.

Our Python algorithm for generating artificially redshifted galaxy images is summarized in the following steps:
(1) We downscaled the size of {\it g}-, {\it r}-, and {\it z}-band DESI images so that the PSF occupy two pixels while preserving their total flux. This is meant to reduce the computation time of step~2, while retaining Nyquist sampling;
(2) For each pixel with S$/$N $\geq3$, we performed a {\it K} correction using the {\tt python} code developed by \cite{Blanton2007}\footnote{https://kcorrect.readthedocs.io/; version 5.0.0 is used.} to calculate the expected flux at the rest-frame wavelength through interpolation. Later, a median ratio of the rest-frame flux derived above to the DESI image flux was calculated and multiplied with the value of pixels with S$/$N $<3$  to obtain a {\it K}-corrected DESI image;
(3) We rescaled the flux using three factors (Eq.~\ref{fscale}). The first one is a dimming factor, caused by the longer luminosity distance at higher redshift. The second one is a brightening factor, caused by the cosmological dilation of wavelength. The first two factors lead to the well-known cosmological dimming. The third factor comes from the monochromatic luminosity evolution, such that the monochromatic luminosity is scaled with $(1+z)^{\alpha}$; 
(4) We downscaled the image size to match the half pixel size and the galaxy angular size as if it appears at high redshifts, while preserving the total flux. In addition to the change of angular size due to longer angular-diameter distances, we took into account the intrinsic galaxy size evolution that is scaled with the physical size $(1+z)^{\beta}$;
(5) We used {\tt photutils} \citep{Bradley2022} to calculate, by Fourier transformation, a kernel that transforms input PSF to the two-time oversampling PSF. Since the input PSF is much smaller, no high frequency noise occurs. The image is convolved with the kernel and was then downscaled in size by 50\% to match the target pixel scale of 0.03\,arcsec/pixel to obtain a resolution-matched image;
(6) We calculated the variance map by multiplying the galaxy flux with the median ratio of variance to flux, generated noise using the map, and added the resulting flux noise to the image. Next, we overlaid the resolution-matched image on top of the clean real background image to produce the final simulated CEERS image.

\begin{figure*}
        \centering
        \includegraphics[width=1\textwidth]{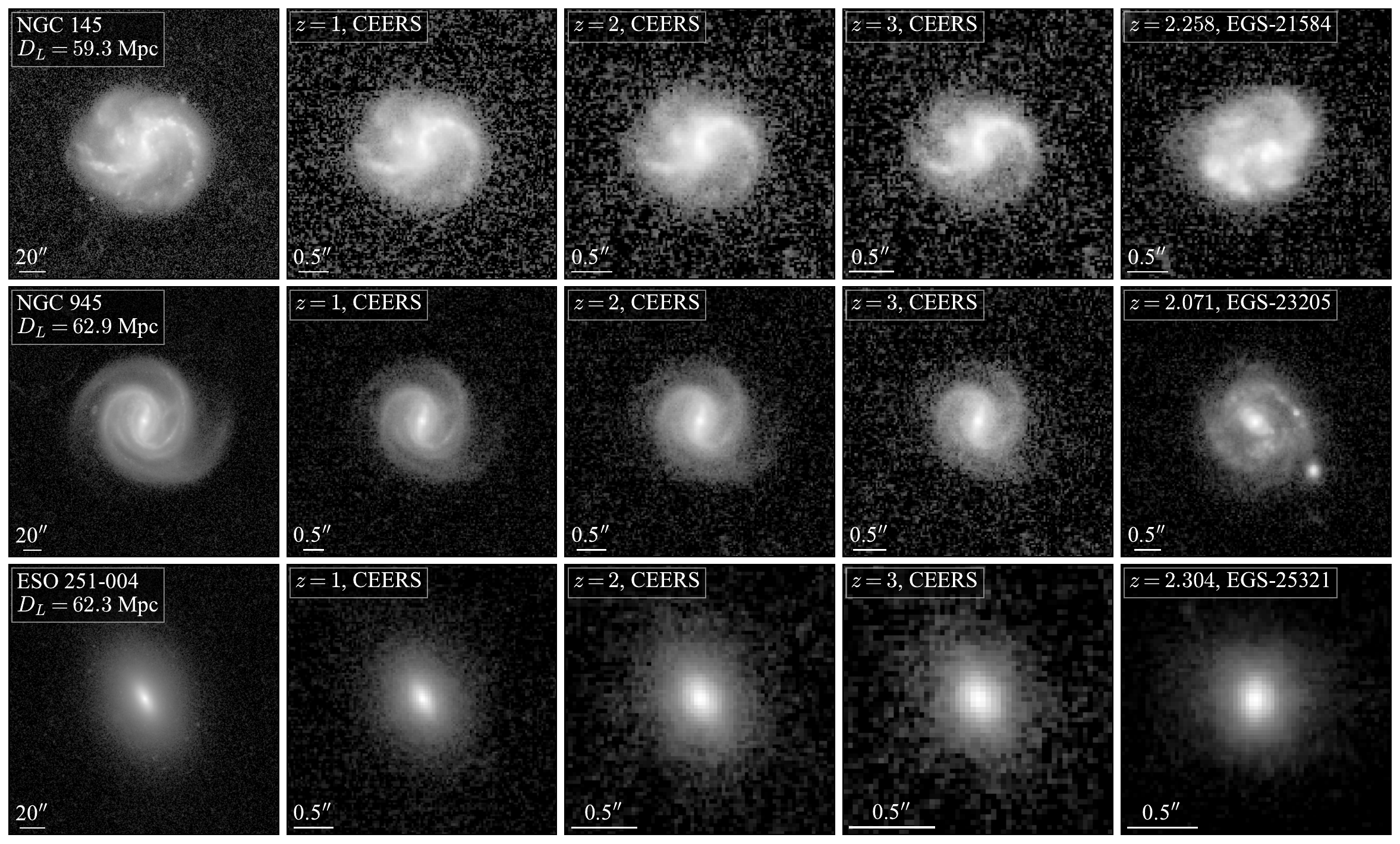}
        \caption{Examples of artificially redshifted galaxy images and the comparison between simulated and real CEERS images. From left to right: First four panels in each row show the DESI {\it r}-band image and the artificial JWST CEERS images at $z=1$, $z=2$, and $z=3$. The fifth column presents real galaxy images in JWST CEERS for comparison. Apparent scales are indicated in each panel. 
        }
        \label{img}
\end{figure*}

We performed the  six steps above for each galaxy to generate its simulated CEERS images at high redshifts. As outlined in step~2, we obtained rest-frame images by interpolating multi-band data at the pixel level, avoiding extrapolation to minimize the risk of introducing significant errors in output flux. To evaluate the uncertainty of this process, we performed a test by using 100 time bootstrap resamplings for one galaxy, which resulted in 100 more simulated images. We found that the resulting uncertainty in the output flux is small, considerably smaller than the typical background noise and galaxy flux noise. Therefore, this source of uncertainty has negligible impact on our results \citep[see also][]{Barden2008}. Figure~\ref{img} illustrates  artificially redshifted galaxy images of three example galaxies. The first column shows the DESI {\it r}-band image, while the  second to forth columns illustrate the artificial JWST CEERS images at $z=1$, $z=2$, and $z=3$.  The final column plots real images of three CEERS galaxies, each with stellar masses comparable to the galaxies in the same row. Although some small-scale structures are suppressed by noise or PSF smoothing effects, the galactic-scale structures such as strong bars and prominent  spirals are still present. Overall, the galaxies remain visible across all redshifts, although they become noisy and blurry. Compared to the simulated CEERS images, real CEERS images of disk looks clumpier. A detailed and rigorous comparison between them is crucial to study the evolution of structure in the future.

\section{Correction of biases and uncertainties} \label{rob}

%
\begin{table}
\renewcommand\arraystretch{1.4}
\caption{Parameters for deriving the correction functions. Results for F115W, F150W, and F200W filters are presented.}            
\label{best-fit}     
\centering                          
\begin{tabular}{c c c c c}        
\hline\hline                
Category & Params & F115W & F150W & F200W \\     
\hline                        
   Size & $\Delta_p$ & 1.14 & 0.91 & 0.77 \\      
   &$\Delta_{50}$    & 0.48 & 0.40 & 0.35 \\
   &$\digamma_1^p$ & 0.238    & 0.229     & 0.225 \\
   &$\digamma_2^p$ & $-0.027$ & $-0.034$  & $-0.043$ \\ 
   &$\digamma_1^{\rm 50,\,cog}$ & 0.226    & 0.221     & 0.223 \\ 
   &$\digamma_2^{\rm 50,\,cog}$ & $-0.036$ & $-0.046$  & $-0.060$ \\ 
   &$\digamma_1^{\rm 50,\,fit}$ & 0.189    & 0.184     & 0.182 \\ 
   &$\digamma_2^{\rm 50,\,fit}$ & $-0.024$ & $-0.025$  & $-0.025$ \\ 
\hline                                   
   Concentration & $g$       & 1.73    & 1.77    & 1.79 \\ 
                 &$\xi_1^b$  & 4.87    & 5.28    & 4.59 \\ 
                 &$\xi_2^b$  & 2.45    & 2.38    & 1.65 \\ 
                 &$\xi_3^b$  & $-0.64$ & $-0.71$ & $-0.71$ \\ 
                 &$\eta_1^C$ & 1.24    & 0.87    & 0.86 \\ 
                 &$\eta_2^C$ & $-0.73$ & $-0.61$ & $-0.67$ \\
\hline                                   
   Asymmetry & $\xi_1^T$ & 31.4    & 27.9    & 7.5 \\ 
             &$\xi_2^T$  & 8.12    & 7.12    & 3.29 \\ 
             &$\xi_3^T$  & $-1.40$ & $-1.43$ & $-1.12$ \\ 
             &$\tau_1$   & 3.79    & 5.33    & 2.96 \\ 
             &$\tau_2$   & $-0.45$ & $-0.72$ & $-0.68$ \\ 
             &$\tau_3$   & 0.57    & 0.54    & 0.51 \\ 
             &$\tau_4$   & 4.97    & 9.35    & 14.98 \\ 
             &$\xi_1^{A_0}$ & 2.28    & 3.59     & 2.06 \\ 
             &$\xi_2^{A_0}$ & 11.47   & 13.05    & 11.04 \\ 
             &$\xi_3^{A_0}$ & $-1.31$ & $-1.48$  & $-1.41$ \\ 
             &$\eta_1^A$    & 0.21    & 0.19     & 0.16 \\ 
             &$\eta_2^A$    & $-0.70$ & $-0.71$  & $-0.69$ \\ 
\hline                                  
Sérsic index & $\eta_1^n$ & 0.092 & 0.089 & 0.085 \\
            & $\eta_2^n$ & $-0.47$ & $-0.47$ & $-0.46$ \\
\hline                                  
Axis ratio & $\eta_1^q$ & 1.17 & 1.05 & 0.99 \\
           & $\eta_1^q$ & $-0.42$ & $-0.39$ & $-0.37$ \\
\hline                                  
\end{tabular}
\end{table}

In this section, we quantify the measurement biases and uncertainties for six commonly used morphological quantities: Petrosian radius, half-light radius, concentration, asymmetry, Sérsic index, and axis ratio. The high-quality, {\it K}-corrected DESI images enable us to accurately determine the intrinsic morphological measurements. We used the $z=2.0$ high-quality {\it K}-corrected DESI images as the training set to derive functions for correcting the biases and uncertainties arising from the resolution effects present in the measurements obtained from degraded images. Using {\it K}-corrected images at other redshifts yields nearly identical results, as the only difference lies in the slight variation in observed rest-frame wavelengths. We reduced their image resolution so that the intrinsic Petrosian radius is $N$ times the PSF FWHM.  We then performed measurements and compared the resulting values with their intrinsic counterparts. We adopted exponentially-growing values of $N$, which are 1.98, 3, 4.55, 6.89, 10.45, 15.83, and 24. The FWHM values for the F115W, F150W, and F200W PSFs are 0.037, 0.049, and 0.064\,arcsec, respectively. These images are denoted as $N$-FWHM images. The typical CEERS noise predominantly impacts the computation of asymmetry. We used the $z=3$ simulated CEERS images, which are the noisiest in our dataset of simulated galaxies, to improve the method for removing noise contribution from the computation of asymmetry.

\subsection{Galaxy size}\label{size}

\begin{figure*}
        \centering
        \includegraphics[width=\textwidth]{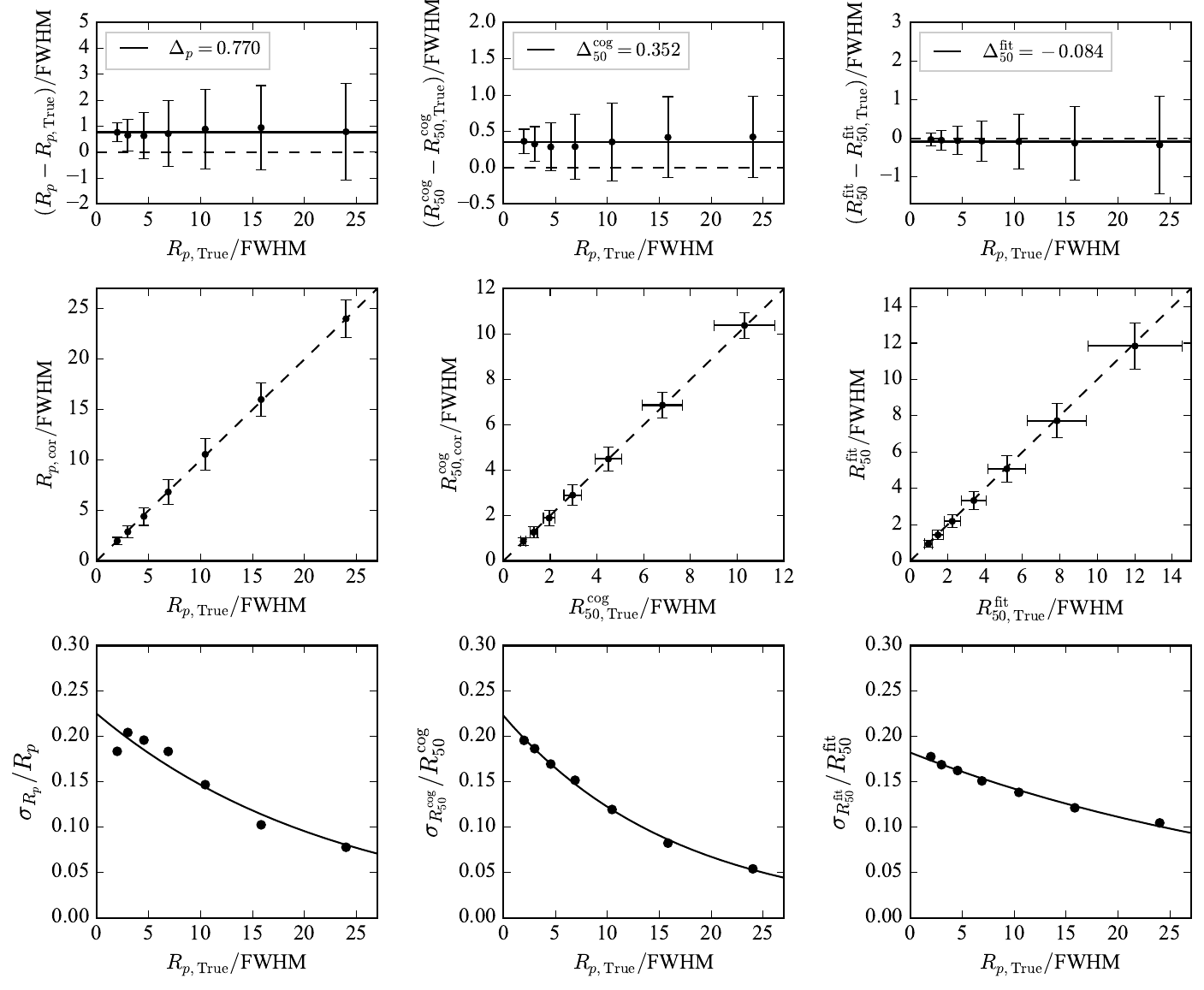}
        \caption{Evaluation of biases and uncertainties in measuring Petrosian radius ($R_p$; left), half-light radius using a non-parametric method ($R^{\rm cog}_{50}$; middle), and half-light radius through Sérsic model fitting  ($R^{\rm fit}_{50}$; right). The top row illustrates the difference between the measured and intrinsic values as a function of $R_{p,\,\rm{True}}/$FWHM, with the solid horizon line marking the mean difference. The middle row displays the correlation between the corrected and intrinsic values, with the one-to-one relation indicated by a dashed line. The bottom row presents the fractional uncertainties, together with their best-fit functions plotted as solid lines.
        }
        \label{Csize}
\end{figure*}

We estimated the flux-weighted center and apparent projection parameters for each image, and measure the Petrosian radius \citep[$R_p$;][]{Petrosian1976}, defined as the radius at which the surface brightness is 20\% of the average surface brightness within $R_p$. We re-measured the center by minimizing galaxy asymmetry (see Section~\ref{Asym}) and used it to re-measure $R_p$ as our final measurement; $R_p$ encompasses at least 99\% of the light within a given galaxy \citep{Bershady2000}.

The half-light radius is another indicator of galaxy size, used to study galaxy evolution \citep[e.g.,][]{Bouwens2004, Buitrago2008}. By measuring the total flux within an elliptical aperture with a radius of $1.5\,R_p$, we derived the fraction of light that the radius enclose as a function of radius, known as the curve of growth (cog). We then determine $R_{20}$, $R_{50}^{\rm cog}$, and $R_{80}$, the radius containing 20\%, 50\%, and 80\% of the total galaxy light, respectively. In additional to the non-parametric approach, we fit a Sérsic model to the galaxy to obtain the half-light radius, denoted as $R_{50}^{\rm fit}$ (see Sect.~\ref{imfit}).

We defined the level of image resolution as $R_{p,\,\rm{True}}/$FWHM. We measured $R_p$, $R_{50}^{\rm cog}$, and $R_{50}^{\rm fit}$ on the F200W $N$-FWHM images and plot the differences between them and their intrinsic values ($R_{p,\,{\rm True}}$, $R_{50,\,{\rm True}}^{\rm cog}$, and $R_{50,\,{\rm True}}^{\rm fit}$)  as a function of $R_{p,\,\rm{True}}/$FWHM in the first row of Fig.~\ref{Csize}. The mean value of $(R_p - R_{p,\,\rm{True}})/$FWHM, denoted by $\Delta_p$, is $0.770$, and that of $(R_{50}^{\rm cog} - R_{\rm 50,\,True}^{\rm cog})/$FWHM, denoted by $\Delta^{\rm cog}_{50}$, is 0.352. Thus, $R_p$ and $R_{50}^{\rm cog}$ are systematically slightly overestimated due to the lower resolution, and the biases should be corrected. Interestingly, the mean difference does not significantly depends on $R_{p,\,\rm{True}}/$FWHM. The detected bias in measuring $R_p$ is consistent with \cite{Whitney2019}, who show that the measured $R_p$ before and after image blurring correlate well with slop of $\sim$\,$1$ and with a systematic offset. In contrast, the mean value of $(R_{50}^{\rm fit} - R_{\rm 50,\,True}^{\rm fit})/$FWHM, denoted by $\Delta^{\rm fit}_{50}$, is so small: $-0.084$,  indicating that the Sérsic fitting can extract the half-light radius without statistical bias, and no correction is needed.

The measurements of $R_p$ and $R_{50}^{\rm cog}$ are corrected for the biases using the functions:
\begin{equation}\label{cor_Rp}
  R_{p,\,\corr}/ \mathrm{FWHM} = R_{p}/ \mathrm{FWHM} - \Delta_p,
\end{equation}
and
\begin{equation}\label{cor_R50}
  R^{\rm cog}_{50,\,\corr}/ \mathrm{FWHM} = R^{\rm cog}_{50}/ \mathrm{FWHM} - \Delta_{50},
\end{equation}
respectively. In Table~\ref{best-fit}, we list $\Delta_p$ and $\Delta_{50}$ for all three filters. The second row of Fig.~\ref{Csize} displays the correlation between the bias-corrected sizes and the intrinsic sizes. No correction was done for $R_{50}^{\rm fit}$. The data points lie closely around the one-to-one relation, marked by a dashed line, indicating that no obvious residual biases exist.

Fractional uncertainty of the size measurement would be more meaningful than the absolute uncertainty, as the measured size is larger in larger galaxies. We calculated the fraction uncertainty as the standard deviation ($\sigma$) of difference between measured and intrinsic values divided by the intrinsic values. We plot the fractional uncertainty as a function of $R_{p,\,\rm{True}}/$FWHM in the third row of Fig.~\ref{Csize}. The size measurement becomes more uncertain at lower resolutions. The following exponential functions, respectively, were fitted to the data: 

\begin{equation}\label{sig_Rp}
  \delta_{R_p}/R_p = \digamma_1^p\cdot \exp(\,\digamma_2^p\cdot x),~{\rm where}~x = R_{p,\,\rm{True}}/{\rm FWHM},
\end{equation}

\begin{equation}\label{sig_R50}
  \delta_{R^{\rm cog}_{50}}/R^{\rm cog}_{50} = \digamma_1^{\rm 50,\,cog}\cdot \exp(\,\digamma_2^{\rm 50,\,cog}\cdot x),~{\rm where}~x = R_{p,\,\rm{True}}/{\rm FWHM},
\end{equation}
and
\begin{equation}\label{sig_R50fit}
  \delta_{R^{\rm fit}_{50}}/R^{\rm fit}_{50} = \digamma_1^{50,\,{\rm fit}}\cdot \exp(\,\digamma_2^{50,\,{\rm fit}}\cdot x),~{\rm where}~x = R_{p,\,\rm{True}}/{\rm FWHM}.
\end{equation}

\noindent
The solid curve in each bottom panel marks the best-fit function. The best-fit parameters for the results based on F115W, F150W, and F200W PSF are listed in Table~\ref{best-fit}.

The functions used to fit the data, as the above functions and those in the rest of this paper, are chosen so that the function can fit the data without obvious residuals and simultaneously meet our expectations regarding the y-axis values when $R_p$ is sufficiently small or large. These functions are empirical. Our primary objective is not to develop a deep understanding of the underlying physics, but rather to describe and characterize the data itself. Empirical functions are used to create a smooth curve that highlights trends or patterns in the data and then used to correct biases and uncertainties, even if the curve itself does not directly represent any physical reasons.

Although our primary focus is on studying the optical morphology of simulated galaxies at $z\leq3$ observed with filters in the 1.15\textendash2.0\,$\mu m$ range, our data set can also shed light on the optical morphology of galaxies at higher redshifts when observed with redder filters, such as F277W, F356W, and F444W. We used a pixel scale of 0.06\,arcsec, created PSFs using {\tt WebbPSF}, and generated {\it N}-FWHM images to carry out the same analysis to comprehend biases and uncertainties involved in quantifying morphology observed through these redder filters. The FWHM values for the F277W, F356W, and F444W PSFs are 0.088, 0.114, and 0.140\,arcsec, respectively. However, we did not perform a Sérsic fitting, as the parameters derived from the fitting exhibited no biases. The parameters for deriving the correction functions are shown in Table~\ref{best-fit2}.

\subsection{Concentration}
\begin{figure*}
        \centering
        \includegraphics[width=0.95\textwidth]{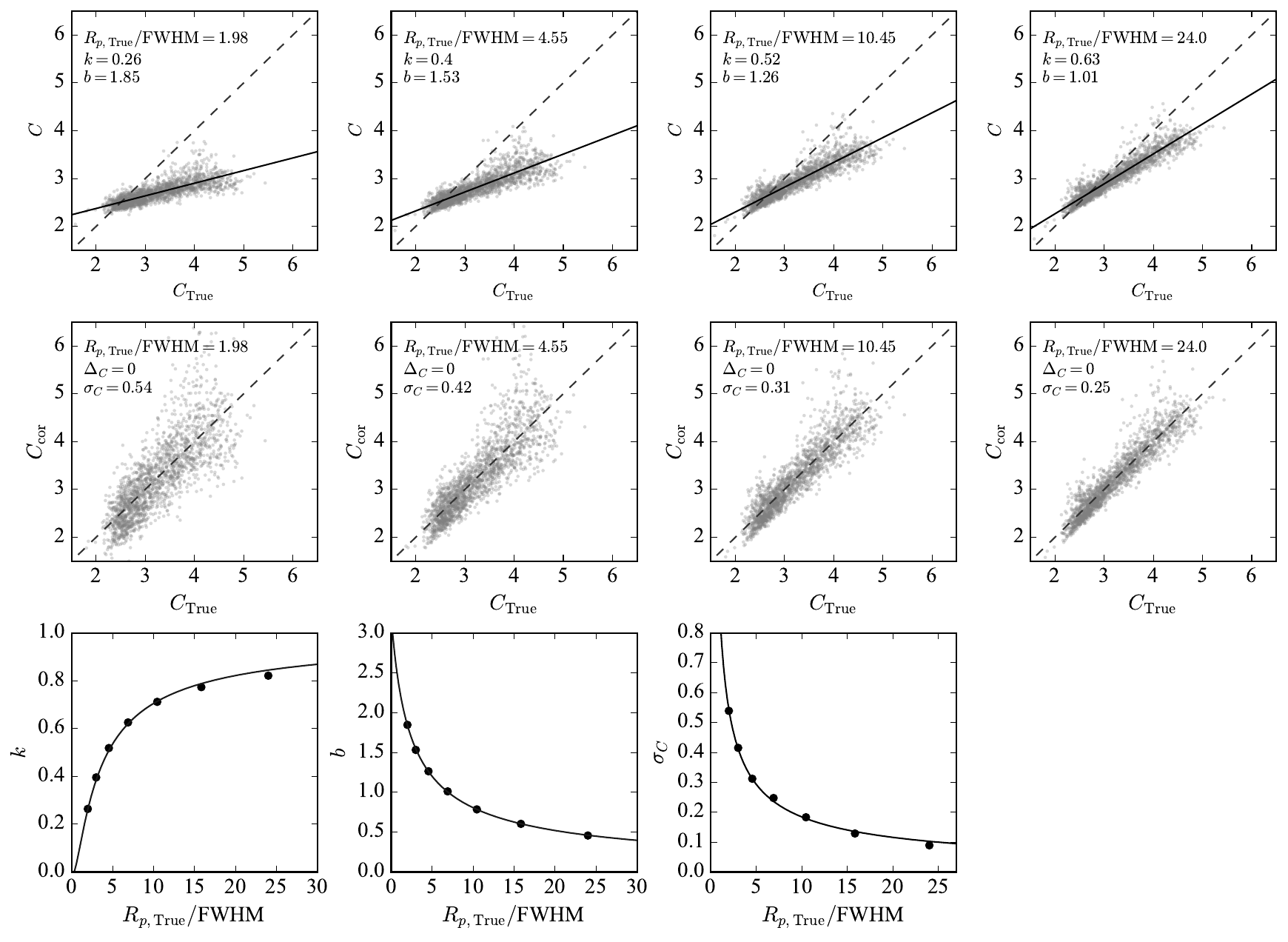}
        \caption{Evaluation of bias and uncertainty in measuring concentration ($C$). Top row shows the correlation between $C$ values obtained from  downsized images ($N$-FWHM images) and their intrinsic values, with the solid line representing the best-fit straight line. Middle row displays the correlation between bias-corrected concentration ($C_{\rm cor}$) and their intrinsic values. Bottom row presents the slop ($k$) and intercept ($b$), and the statistical uncertainty ($\sigma_C$), as a function of $R_{p,\,\rm{True}}/$FWHM, with the curve marking their best-fit functions. 
        }
        \label{CCon}
\end{figure*}

The concentration ($C$) measures the degree to which a galaxy's light distribution is centrally concentrated. Following the definition in \cite{Conselice2003}, we compute $C$ as
\begin{equation}
  C=5\cdot\log_{10}\left(\frac{R_{80}}{R_{20}}\right).
\end{equation}

\noindent
Higher values indicate more centralized light distributions. We denote intrinsic concentration measured in the {\it K}-corrected DESI image as $C_{\rm True}$. In the first row of Fig.~\ref{CCon}, we plot $C$ measured from the $N$-FWHM images against $C_{\rm True}$. The first, second, third, and final columns display the results of image resolution levels $R_{p,\,\rm{True}}/{\rm FWHM}=1.98$, 4.55, 10.45, and 24, respectively. The results show that $C$ is systematically underestimated, with a more significant underestimation for galaxies with higher $C_{\rm True}$. This effect is more pronounced in angularly smaller galaxies with lower resolution (lower $R_{p,\,\rm{True}}/$FWHM). The correlations between $C$ and $C_{\rm True}$ are nearly linear and, hence, we fit the data with a linear function:

\begin{equation}
  C = k \cdot C_{\rm True} + b
.\end{equation}
\noindent
The best-fit parameters are listed at the top of each panel in the first row and are plotted as a function of $R_{p,\,\rm{True}}/$FWHM in the bottom row. Toward lower resolutions, the best-fit slop $k$ decreases, while the best-fit intercept $b$ increases. The underestimation of measured $C$ is attributed to the PSF smoothing effect, which overestimates $R_{20}$ more than $R_{80}$. Our findings are consistent with \cite{Whitney2021} and \cite{Yeom2017}, who showed that $C$ values are more underestimated at higher redshift, where galaxies are smaller and resolutions are lower. Nevertheless, the literature still lacks an accurate correction function to address the issue, which will be further explored in the following discussion.

We use the best-fit straight lines to correct for the bias and uncertainty in measuring $C$. The correction function is given by:
\begin{equation}\label{cor_C}
  C_{\rm cor} = (C - b)/k.
\end{equation}

\noindent
The correlations between $C_{\rm cor}$ and $C_{\rm True}$ are shown in the middle row of Fig.~\ref{CCon}. The mean value ($\Delta_C$) and standard deviation ($\sigma_C$) of the difference between $C_{\rm cor}$ and $C_{\rm True}$ are presented at the top of each panel in the middle row. Also, $\Delta_C$ are zero at all examined image resolution levels. $\sigma_C$ serves as the statistical measurement uncertainty, which increases with lower resolution. The large uncertainty at low resolution stems from the flatten relationship between $C$ and $C_{\rm Truen}$ or, in other words, the low $k$ value. $\sigma_C$ as a function of $R_{p,\,\rm{True}}/$FWHM is plotted in the bottom row.

In practical applications, we observe galaxies of various sizes, which necessitates the development of functions to derive the $k$ and $b$ for a given resolution level. To achieve this goal, we fit $k$ versus $R_{p,\,\rm{True}}/$FWHM using the following function:

\begin{equation}\label{corr_k}
  k = \frac{-2\cdot(x+1)}{1+(x+1)^g} + 1,~{\rm where}~x = R_{p,\,\rm{True}}/{\rm FWHM}.
\end{equation}

\noindent 
This function converges to 1 when $R_{p,\,\rm{True}}$ is sufficiently large. We fit $b$ versus $R_{p,\,\rm{True}}/$FWHM using the following function:

\begin{equation}\label{corr_b}
  b = \xi_1^{\,b}\cdot(x+\xi_2^{\,b})^{\,\xi_3^{\,b}},~{\rm where}~x = R_{p,\,\rm{True}}/{\rm FWHM}.
\end{equation}

\noindent 
This function converges to 0 when $R_{p,\,\rm{True}}$ is sufficiently large. 
Therefore, for a given $R_{p,\,\rm{True}}$, the correction function can be derived using Eq.~(\ref{cor_C}), (\ref{corr_k}), and (\ref{corr_b}). 
To estimate the statistical uncertainty for a given resolution level, we fit $\sigma_C$ versus $R_{p,\,\rm{True}}/$FWHM using the following function:

\begin{equation}\label{sig_C}
  \sigma_C = \eta_1^C \cdot x^{\,\eta_2^C},~{\rm where}~x = R_{p,\,\rm{True}}/{\rm FWHM}.
\end{equation}

\noindent
The best-fit functions are plotted as solid curves. The best-fit parameters $g$, $\xi_1^{\,b}$, $\xi_2^{\,b}$, $\xi_3^{\,b}$, $\eta_1^C$, and $\eta_2^C$  for the results based on F115W, F150W, and F200W PSF are listed in Table~\ref{best-fit}. Those based on F277W, F356W, and F444W are provided in Table~\ref{best-fit2}.

\subsection{Asymmetry} \label{Asym}

\begin{figure*}
        \centering
        \includegraphics[width=1\textwidth]{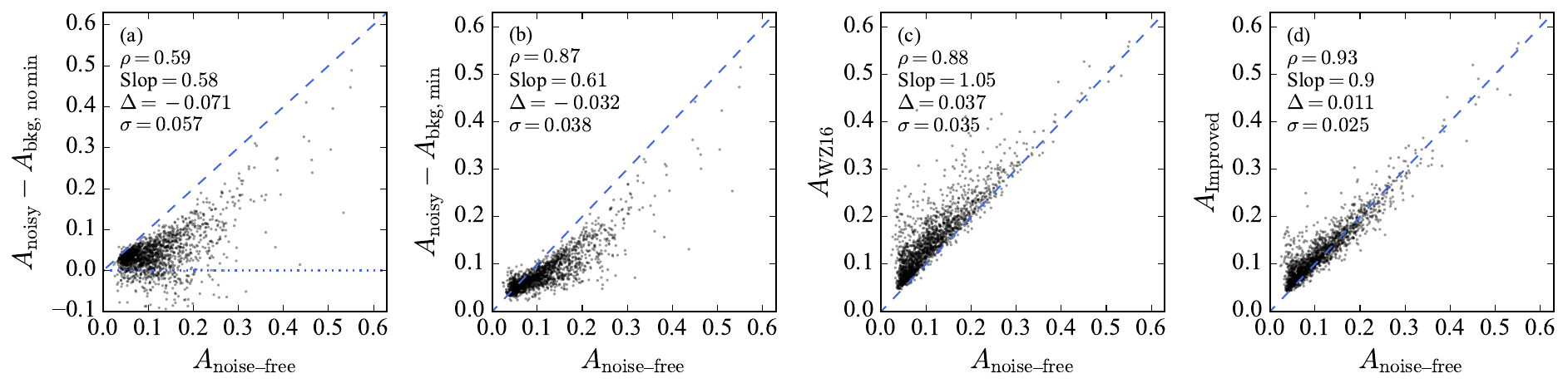}
        \caption{Comparison of different noise correction methods applied to asymmetry. The four panels present the results for (a) calculating background asymmetry without minimization ({\tt statmorph} is used), (b) calculating background asymmetry with minimization, (c) Eq.~(\ref{WZ}) with $f_1=1$ and $f_2=\sqrt{2}$, proposed by \cite{Wen2016}, and (d) Eq.~(\ref{WZ}) with $f_1=2.25$ and $f_2=2.1$, our proposed values. Our improved noise correction exhibits enhanced performance in correcting for noise effects and reproducing the noise-free asymmetry.
        }
        \label{CAsky}
\end{figure*}
\begin{figure*}
        \centering
        \includegraphics[width=0.95\textwidth]{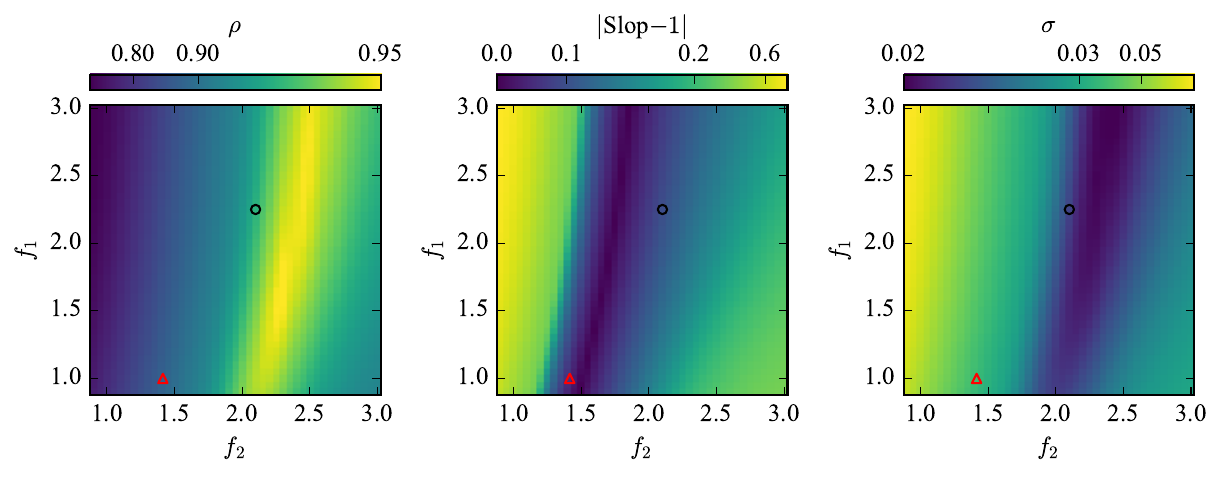}
        \caption{Pearson correlation coefficient ($\rho$), absolute difference between the slop and unity ($|{\rm slop}-1|$), and scatter ($\sigma$) between asymmetry calculated using Eq.~(\ref{WZ}) and $A_{\rm noise-free}$ as a function of $f_1$ and $f_2$. The values are color-encoded using histogram equalization scale. The black circle marks the optimal $f_1=2.25$ and $f_2=2.1$ we determine empirically, while the red triangle marks the $f_1=1$ and $f_2=\sqrt{2,}$ as suggested in \cite{Wen2016}.
        }
        \label{f12}
\end{figure*}

\subsubsection{Definition}

The asymmetry ($A$) measures the degree to which a galaxy's light distribution is 180$\degr$rotationally symmetric.  Originally, $A$ is determined by rotating the galaxy image by 180$\degr$ about the galaxy center, set as the pixel location of the maximum value, and subtracting the rotated image from the galaxy image to obtain a difference image. Then, $A$ is calculated as 0.5 times the ratio of the summation of the absolute pixel values in the difference image to the summation of the absolute pixel values in the original image \citep{Abraham1996}. The background asymmetry is calculated in the same vein for a portion of sky and subtracted.  To solve the problem of substantial variation in $A$ due to uncertainty in the center determination, the algorithm is improved by including a process for searching for a new galaxy center to minimize the asymmetry; background asymmetry is also minimized in the same vein and then subtracted \citep{Conselice2000, Lotz2004}. In addition, the factor of 0.5 in the original definition is dropped. This is the most conventional method to calculate galaxy asymmetry. The calculation is given by:
\begin{equation}\label{asymm}
\begin{split}
  A_{\rm C00} & = \frac{\min(\sum |I_0-I_{180}|)}{\sum |I_0|} - \frac{\min(\sum |B_0-B_{180}|)}{\sum |I_0|} \\
  & = A_{\rm noisy} - A_{\rm bkg,\,min},
\end{split}
\end{equation}

\noindent
where $I$ represents the galaxy image, and $B$ represents the sky background.  The subscript of $A_{\rm C00}$ marks \cite{Conselice2000}. The summation is done over all pixels within a 1.5\,$R_p$ elliptical aperture, which has the apparent projection parameters determined in Sect.~\ref{size}, centered on the galaxy.

Although the minimization of background asymmetry has been proposed for two decades, there are some studies stick to the background asymmetry without minimization, calculated as follows:

\begin{equation}
  A_{\rm bkg,\,no\,min} = \frac{\sum |B_0-B_{180}|}{\sum |I_0|}. 
\end{equation}

\noindent
For example, \cite{Rodriguez-Gomez2019} developed the PYTHON package {\tt statmorph} to quantify galaxy morphology and calculate background asymmetry without minimization within a sky box located beside the galaxy segmentation.   Similarly, \cite{Tohill2021} constructed a 10-pixel by 10-pixel grid over the image area outsides the galaxy segmentation to perform the calculation.

However, it has been already shown that $A_{\rm bkg,\,min}$ overestimates the contribution of noise to the calculation of galaxy asymmetry \citep{Shi2009, Wen2016}, and so the $A_{\rm bkg,\,no\,min}$ even more severely overestimates the noise contribution. \cite{Shi2009} showed that asymmetry values of galaxies in the field of shallower observations are systematically lower than those of the same galaxies in the field of deeper observations. These authors proposed  measuring the distribution of noise asymmetries in randomly selected regions surrounding the target galaxy and calculate the 15\% probability low-end tail as the final background asymmetry measurement. Although this method could remove the bias on average, but significant scatter still persists.\cite{Wen2016} studied asymmetries of galaxies before and after making them noisy and proposed a new noise correction, which is given by:

\begin{equation}\label{WZ}
  A = \frac{\min(\sum|I_0-I_{180}|) - F_2\cdot\min(\sum|B_0-B_{180}|)   }{\sum|I_0| - F_1\cdot \sum|B_0|}, 
\end{equation}
\noindent
where
\begin{equation}
  F_1 = \frac{N_{I_0\,<\,f_1\cdot \sigma_{\rm bkg}}}{N_{\rm all}}
,\end{equation}
\noindent
and 
\noindent
\begin{equation}
  F_2 = \frac{N_{|I_0-I_{180}|\,<\,f_2\cdot \sigma_{\rm bkg}}}{N_{\rm all}}, 
\end{equation}

\noindent
where $N_{\rm all}$ represents the number of pixels encloses by the 1.5\,$R_p$ elliptical aperture.  $N_{I_0\,<\,f_1\cdot \sigma_{\rm bkg}}$ is the number of pixels  dominated by noise in the galaxy image, selected as those with values less than $f_1\cdot \sigma_{\rm bkg}$; $N_{|I_0-I_{180}|\,<\,f_2\cdot \sigma_{\rm bkg}}$ is the number of pixels dominated by noise in the difference image, selected as those with values less than $f_2\cdot \sigma_{\rm bkg}$. This correction is based on the fact that only noisy pixels are affected. \cite{Wen2016} suggested using $f_1=1$ and $f_2=\sqrt{2}$, which, however, are not the optimal choices, as discussed below.

\subsubsection{Improved noise correction}\label{ImpNC}

To understand the noise contribution, we measured the asymmetry for the resolution-match images (simulated noise not added yet), which are obtained from step~5 of the redshifting procedure and have such high S$/$N that the noise contribution is almost negligible. We denote the result as $A_{\rm noise-free}$. We note that $A_{\rm noise-free}$ is still biased due to resolution effects and will be addressed later. We then measured the asymmetry for the  simulated CEERS images using {\tt statmorph}, in which no minimization of the background asymmetry was done, and we denote the result as $A_{\rm noisy}-A_{\rm bkg,\,no\,min}$. We performed the measurement with minimization using Eq.~(\ref{asymm}) and denote the result as $A_{\rm noisy}-A_{\rm bkg,\,min}$. We performed the measurement using Eq.~(\ref{WZ}), adopting $f_1=1$ and $f_2=\sqrt{2}$, and denote the result as $A_{\rm WZ16}$. We plot these as a function of $A_{\rm noise-free}$ in Fig.~\ref{CAsky}.

In a perfect noise correction, $A_{\rm noise-free}$ should be accurately reproduced. We first confirm in the panel (a) that $A_{\rm bkg,\,no\,min}$ severely overestimates noise contribution, leading to an underestimation of asymmetry and even physically meaningless negative values. As shown in panel (b), $A_{\rm bkg,\,min}$ works better than $A_{\rm bkg,\,no\,min}$, but the overestimation of noise contribution is still significant, especially at higher $A_{\rm noise-free}$ values. The mean difference between $A_{\rm noisy}-A_{\rm bkg,\,min}$ and $A_{\rm noise-free}$ gives $-0.032$. This correlation is sublinear. Panel (c) shows that $A_{\rm WZ16}$ underestimate noise contribution by $0.037$, while the main improvement is that the correlation between $A_{\rm WZ16}$ and $A_{\rm noise-free}$ is brought close to linear.

The problem related to the underestimation of noise contribution in using Eq.~(\ref{WZ}) may be solved if a larger fraction of pixels are defined as noisy pixels. We searched for an optimal solution by testing different value of $f_1$ and $f_2$. We calculated the asymmetry using Eq.~(\ref{WZ}) by adopting $f_1=0.9$, 0.95, 0.1, ..., 3.0 and $f_2=0.9$, 0.95, 0.1, ..., 3.0. For each pair of $f_1$ and $f_2$, we calculated the Pearson correlation coefficient, slop, and scatter between the resulting asymmetry and $A_{\rm noise-free}$. The coefficient, $|{\rm slop}-1|$, and scatter as a function of $f_1$ and $f_2$ are plotted in Fig.~\ref{f12}.

The results show that the coefficient and scatter reach their maximum and minimum values at $f_2\approx2.3$, while the slop reaches 1 at $f_2\approx1.7$. The values suggested by \cite{Wen2016} are marked by a red triangle, which would get a nearly linear relation, but the relation will be dispersive. Constructing the tightest relation will make the relation sublinear. It is thus not feasible to use Eq.~(\ref{WZ}) to obtain corrected asymmetry that has a relationship with $A_{\rm noise-free}$ that is very tight and linear simultaneously. As a compromise, we derive the optimal $f_1$ and $f_2$ by requiring that the slop is greater than 0.9 and less than 1.1, and the scatter achieves the minimum value, resulting in $f_1=2.25$ and $f_2=2.1$. Minor changes in the criterion do not significantly impact our results. The optimal choice is marked by black circle in Fig.~\ref{f12}. These values are used to derive noise-removed asymmetry, denoted as $A_{\rm Improve}$, which is plotted against $A_{\rm noise-free}$ in the panel (d) of Fig.~\ref{CAsky}. It is shown that the correlation between $A_{\rm Improve}$ and $A_{\rm noise-free}$ is tight ($\rho=0.93$), almost linear (slop $=0.9$), and has a small residual bias ($\Delta=0.011$) and small scatter ($\sigma=0.025$). The scatter is also smaller than the result obtained by the noise correction proposed by \cite{Shi2009}, which yield 0.044 (see their Fig.~15). Our improved noise correction exhibits enhanced performance in correcting for noise effects and reproducing the noise-free asymmetry. Nevertheless, we would like to point out that there is still a minor underestimation of noise contribution present, that is, an overestimation of asymmetry, when $A_{\rm noise-free}\lesssim 0.1$.

\subsubsection{Correction of resolution effects}

\begin{figure*}
        \centering
        \includegraphics[width=0.95\textwidth]{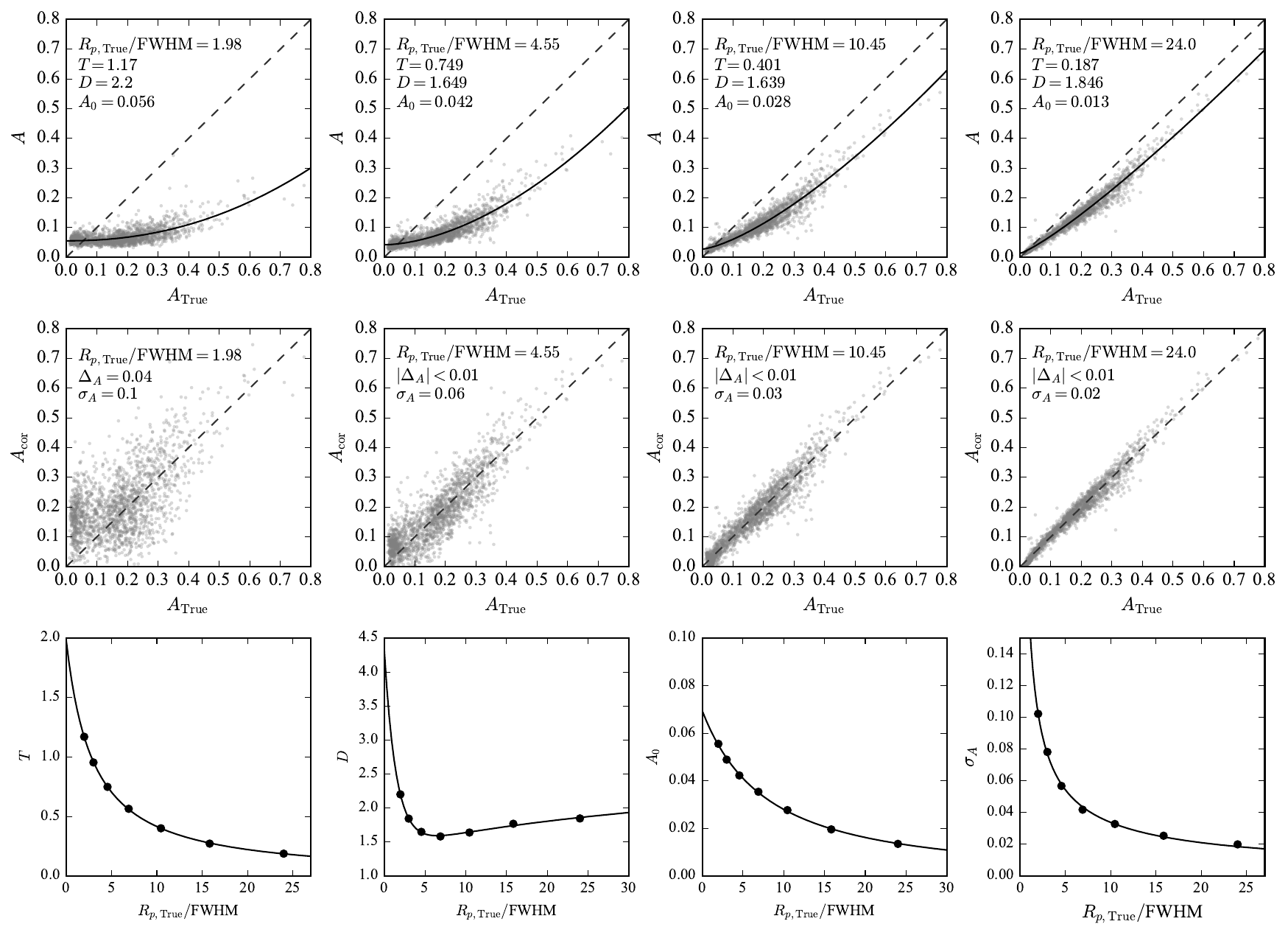}
        \caption{Evaluation of bias and uncertainty due to resolution effects in measuring asymmetry ($A$). Top row shows the correlation between $A$ values obtained from the $N$-FWHM images and their intrinsic values ($A_{\rm True}$), with the solid curve representing the best-fit function. Middle row displays the correlation between bias-corrected asymmetries ($A_{\rm cor}$) and $A_{\rm True}$. Bottom row presents the parameters ($T$, $D$, and $A_0$) and the uncertainty in correcting $A$ ($\sigma_A$), as a function of $R_{p,\,\rm{True}}/$FWHM, with the curve marking their best-fit functions.
        }
        \label{CApsf}
\end{figure*}

In addition to the noise contribution, the resolution effect also plays a significant role in affecting asymmetry measurements. This effect becomes particularly crucial when studying high-redshift galaxies, which are less spatially resolved.  Focusing on changes in the apparent galaxy size relative to a fixed pixel size, \cite{Conselice2000} demonstrated that measured galaxy asymmetry is increasingly reduced, as the apparent size of a galaxy decreases. We measured galaxy asymmetries using re-binned images without PSF convolution, and found that the binning effect is negligible compared to the PSF effect. It has also been found that galaxy asymmetries are underestimated at higher redshifts \citep{Conselice2003, Yeom2017, Whitney2021}; however, this should be partly attributed to the overestimation of noise contribution using the conventional method, as discussed in Sect.~\ref{ImpNC}, and partly to resolution effects. An accurate correction method for biases caused by resolution effects has not been developed yet.

\begin{figure}
        \centering
        \includegraphics[width=0.5\textwidth]{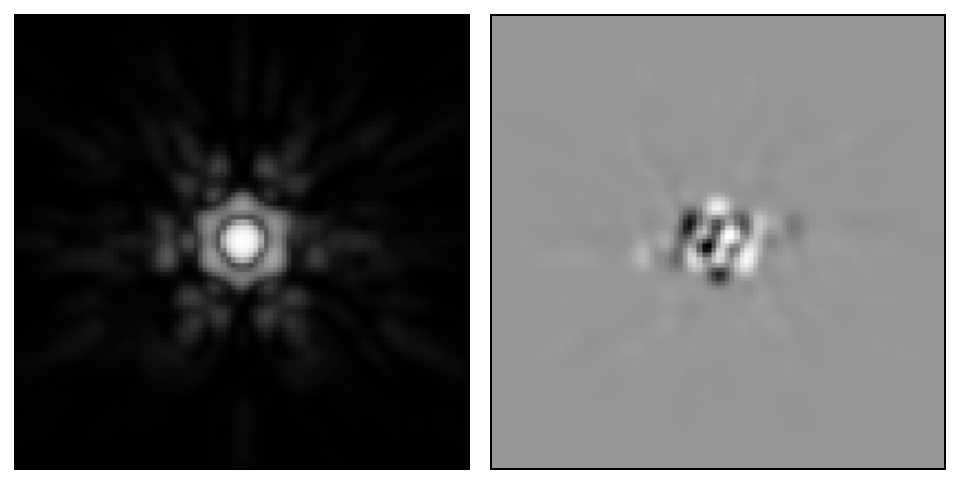}
        \caption{Illustration of the asymmetry of two-time oversampling JWST F200W PSF. The PSF is plotted in logarithmic scale on the left, while the difference between the PSF and its 180$\degr$-rotational image is plotted in linear scale on the right. 
        }
        \label{Apsf}
\end{figure}

To account for resolution effects, we first measured intrinsic asymmetries from the high-quality, {\it K}-corrected DESI images and denote them as $A_{\rm True}$.  We then measured $A$ from the {\it N}-FWHM images and present $A$ as a function of $A_{\rm True}$ in the top row of Fig.~\ref{CApsf}. The first, second, third, and final columns display results obtained at the image resolution levels $R_{p,\,{\rm True}}/{\rm FWHM}=1.98$, 4.55, 10.45, and 24, respectively. The results show that when galaxies are the most intrinsically symmetric ($A_{\rm True}\lesssim0.05$) and are at low resolution ($R_{p,\,{\rm True}}/{\rm FWHM}\lesssim4.55$), the measured $A$ slightly overestimates $A_{\rm True}$. This overestimation is attributed to the asymmetry of the {\it JWST} PSF.  Figure~\ref{Apsf} displays the two-time oversampling F200W PSF on the left and the difference between the PSF and its 180$\degr$-rotational image on the right. The PSF asymmetry causes the convolution to make intrinsically symmetric galaxies appear asymmetric.  In contrast, when galaxies are intrinsically asymmetric ($A_{\rm True}\gtrsim 0.05$), the measured $A$ values are underestimated, with a greater extent at higher $A_{\rm True}$.  This underestimation is due to PSF convolution smoothing out asymmetric structures, particularly the most asymmetric ones. The smoothing effect is more efficient than the effect caused by PSF asymmetry in asymmetric galaxies.

The $A_{\rm True}$\textendash$A$ relations are nonlinear, especially at low resolutions. We fit the data at each resolution level with the following function:

\begin{equation}\label{Afit}
  A = A_{\rm True}\cdot (A_{\rm True}/D)^{T} + A_0,
\end{equation}

\noindent
where $D$ and $T$ control the flatness of the function and $A_0$ is the y-intercept representing the asymmetry of a fully symmetric galaxy after convolving a asymmetric PSF. The best-fit functions are plotted as black solid curves. We reverse Eq.~(\ref{Afit}) to estimate the intrinsic value for a given measured $A$ value. However, this operation is only available for $A\geq A_0$. To estimate the intrinsic value for $A< A_0$, we replace $A$ with $2A-A_0$, which is the symmetric value of $A$ with respect to $y=A_0$, to do the estimation. The correction function is as follows:

\begin{equation}\label{Acorf}
  A_{\rm cor} = D\cdot\left(\frac{\Delta A}{D}\right)^{\frac{1}{1+T}},~{\rm  where}~ \Delta A = 
  \begin{cases}
        A - A_0,~~~~{\rm if}~A \geq A_0\\
        2A_0 - A,~~{\rm if}~A< A_0 \\
  \end{cases}
  .
\end{equation}

\noindent
The second row in Fig.~\ref{CApsf} plots the correlations between $A_{\rm cor}$ and $A_{\rm True}$. The mean difference ($\Delta_A$) and scatter ($\sigma_A$) between them are presented at the top of each panel. 
At high resolution, 
$\Delta_A$ is small ($|\Delta_A|$\,$<$\,$0.01$). However, 
At low resolution, a non-negligible $\Delta_A$ value ($\sim$\,$0.04$) is observed, indicating that a residual small bias still exists when the resolution is sufficiently low.  Meanwhile, $\sigma_A$ is significant ($\sim$\,$0.1$) at $R_{p,\,{\rm True}}/{\rm FWHM}<4.55$, which stems from the flatness of the $A_{\rm True}$\textendash$A$ relation.

To obtain the correction function for a given resolution level, we study the parameters $T$, $D$, and $A_0$ as a function of $R_{p,\,{\rm True}}/$FWHM in the bottom row of Fig.~\ref{CApsf}. We respectively fit the correlations between $T$, $D$, and $A_0$ with $R_{p,\,{\rm True}}/$FWHM using the functions: 

\begin{equation}\label{T}
  T = \xi_1^T\cdot(x+\xi_2^T)^{\,\xi_3^T},~{\rm where}~x = R_{p,\,{\rm True}}/{\rm FWHM},
\end{equation}

\begin{equation}\label{D}
  D = \tau_1 \cdot\exp(\tau_2\cdot x)+\tau_3 \cdot \ln (x+\tau_4),~{\rm where}~x = R_{p,\,{\rm True}}/{\rm FWHM},
\end{equation}

\noindent
and

\begin{equation}\label{A0}
  A_0 = \xi_1^{A_0}\cdot(x+\xi_2^{A_0})^{\,\xi_3^{A_0}},~{\rm where}~x = R_{p,\,{\rm True}}/{\rm FWHM}.
\end{equation}

\noindent
The best-fit functions are marked with black solid curves. The correction function Eq.~(\ref{Acorf}) can therefore be derived for a given galaxy size through Eq.~(\ref{T})\textendash(\ref{A0}). To estimate the statistical uncertainty for a given resolution level, we plot $\sigma_A$ against $R_{p,\,{\rm True}}/$FWHM in Fig.~\ref{CApsf} and fit them using the following function:

\begin{equation}\label{sig_A}
  \sigma_A = \eta_1^A \cdot x^{\,\eta_2^A},~{\rm where}~x = R_{p,\,{\rm True}}/{\rm FWHM}.
\end{equation}
The best-fit function is plotted as a black solid curve. The best-fit parameters $\xi_1^T$, $\xi_2^T$, $\xi_3^T$, $\tau_1$, $\tau_2$, $\tau_3$, $\tau_4$, $\xi_1^{A_0}$, $\xi_2^{A_0}$, $\xi_3^{A_0}$, $\eta_1^A$, and $\eta_2^A$ for the results based on F115W, F150W, and F200W PSF are listed in Table~\ref{best-fit}.  Those based on F277W, F356W, and F444W are shown in Table~\ref{best-fit2}.

\begin{table}
\renewcommand\arraystretch{1.4}
\caption{Parameters for deriving the correction functions. Results for F277W, F356W, and F444W filters are presented.}        
\label{best-fit2}     
\centering                         
\begin{tabular}{c c c c c}       
\hline\hline                
Category & Params & F277W & F356W & F444W \\    
\hline                      
   Size & $\Delta_p$ & 0.73 & 0.69 & 0.65 \\    
   &$\Delta_{50}$ & 0.33 & 0.32    & 0.31 \\
   &$\digamma_1^p$ & 0.210 & 0.211    & 0.213 \\
   &$\digamma_2^p$ & $-0.035$ & $-0.042$    & $-0.047$ \\ 
   &$\digamma_1^{\rm 50,\,cog}$ & 0.209 & 0.215    & 0.218 \\ 
   &$\digamma_2^{\rm 50,\,cog}$ & $-0.049$ & $-0.061$    & $-0.070$ \\ 
\hline                                  
   Concentration & $g$ & 1.78 & 1.79    & 1.81 \\ 
                 &$\xi_1^b$ & 8.90 & 5.83    &  5.00 \\ 
                 &$\xi_2^b$ & 3.91 & 2.23    & 1.70 \\ 
                 &$\xi_3^b$ & $-0.90$ & $-0.80$    & $-0.77$ \\ 
                 &$\eta_1^C$ & 0.97 & 0.85    & 0.85 \\ 
                 &$\eta_2^C$ & $-0.71$ & $-0.68$    & $-0.71$ \\ 
\hline                                 
   Asymmetry &$\xi_1^T$ & 51.3    & 10.7    & 5.5 \\
             &$\xi_2^T$ & 9.46    & 4.78    & 3.05 \\
             &$\xi_3^T$ & $-1.61$ & $-1.23$ & $-1.07$ \\ 
             &$\tau_1$  & 5.73    & 3.20    & 2.29 \\ 
             &$\tau_2$  & $-0.81$ & $-0.68$ & $-0.67$ \\ 
             &$\tau_3$  & 0.49    & 0.49    & 0.46 \\ 
             &$\tau_4$  & 18.5    & 20.6    & 34.1 \\ 
             &$\xi_1^{A_0}$ & 1612.6  & 384.0   & 580.2 \\
             &$\xi_2^{A_0}$ & 36.7    & 31.3    & 33.4 \\ 
             &$\xi_3^{A_0}$ & $-2.83$ & $-2.59$ & $-2.70$ \\ 
             &$\eta_1^A$ & 0.20    & 0.16    & 0.14 \\ 
             &$\eta_2^A$ & $-0.75$ & $-0.70$ & $-0.66$ \\ 
\hline                                 
\end{tabular}
\end{table}

\subsection{Sérsic index and axis ratio} \label{imfit}

We use {\tt IMFIT} \citep{Erwin2015} to carry out a two-dimensional (2D) fitting using a single Sérsic model \citep{Sersic1968} on the images. This allows us to determine the half-light radius ($R_{50}^{\rm fit}$), Sérsic index ($n$), and axis ratio ($q$). We use a two-time oversampling PSFs in our analysis. {\tt IMFIT} finds the optimal model by adjusting the 2D function parameters through nonlinear minimization of total $\chi^2$.  The Levenberg-Marquardt algorithm is used for the $\chi^2$ minimization. We set the lower and upper bounds of $n$ for the fitting to 0.5 and 6, respectively. The bias and uncertainty in measuring $R_{50}^{\rm fit}$ have been discussed in Sect.~\ref{size}.

\begin{figure}
        \centering
        \includegraphics[width=0.45\textwidth]{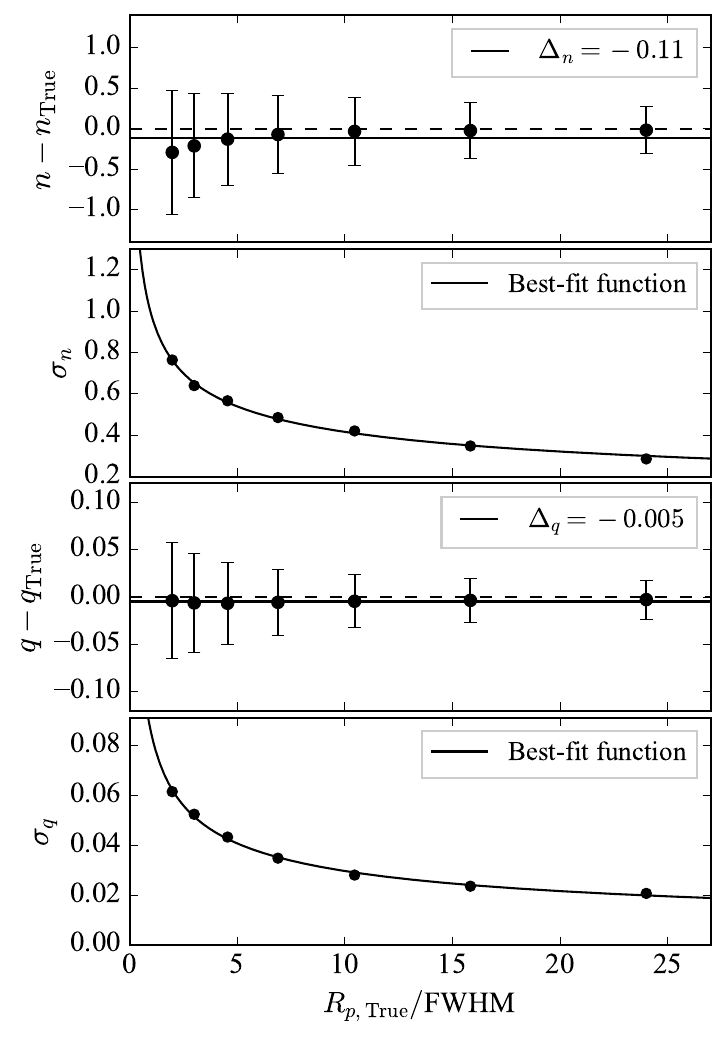} 
        \caption{Evaluation of biases and uncertainties due to resolution effects in measuring Sésic index ($n$) and axis ratio ($q$) using {\tt IMFIT}.
        }
        \label{QNpsf}
\end{figure}

We denote intrinsic $n$ and $q$ values measured from the high-quality, {\it K}-corrected DESI images as $n_{\rm True}$ and $q_{\rm True}$, respectively. We plot the difference between $n$ measured from the {\it N}-FWHM image and $n_{\rm True}$, that is $n-n_{\rm True}$, as a function of $R_{p,\,\rm{True}}/$FWHM in the first row of Fig.~\ref{QNpsf}. The mean difference, $\Delta_n=-0.11$, indicates that the impact of resolution degradation is small, properly because {\tt IMFIT} already accounts for PSF effects.  As we go on to show in Sect.~\ref{validation}, there is no obvious difference on average between intrinsic values and those measured on the simulated CEERS images. Our findings are in line with previous studies based on model galaxies (\citealt{Barden2008}; \citealt{Davari2016}; \citealt{Euclid2023XXVI}; but see \citealt{Paulino-Afonso2017}). We thus refrain from developing a correction function for $n$. The second row plots the profile of measurement uncertainty $\sigma_n$, which we fit with the function:

\begin{equation}\label{sig_n}
  \sigma_n = \eta_1^n \cdot x^{\,\eta_2^n},~{\rm where}~x = R_{p,\,{\rm True}}/{\rm FWHM}. 
\end{equation}

\noindent
The black curve represents the best-fit function. The measurement of $n$ becomes more uncertain with lower resolutions.

Similarly, we calculate $q-q_{\rm True}$ and plot it as a function of $R_{p,\,\rm{True}}/$FWHM in the third row of Fig.~\ref{QNpsf}. The mean difference is very small ($\Delta_q=-0.005$), indicating little or no measurement bias.  The bottom panel presents the profile of statistical uncertainty $\sigma_q$, which we fit with the function:

\begin{equation}\label{sig_q}
  \sigma_q = \eta_1^q \cdot x^{\,\eta_2^q},~{\rm where}~x = R_{p,\,{\rm True}}/{\rm FWHM}. 
\end{equation}

\noindent
The black curve marks the best-fit function. The measurement of $q$ becomes more uncertain at lower resolutions, yet these uncertainties remain negligible when compared to the broad dynamical range of $q$. The best-fit parameters $\eta_1^n$, $\eta_2^n$, $\eta_1^q$, and $\eta_2^q$ for the results based on F115W, F150W, and F200W PSFs are provided in Table~\ref{best-fit}.

\section{Application to simulated CEERS images}\label{validation}
\begin{figure*}
        \centering
        \includegraphics[width=0.95\textwidth]{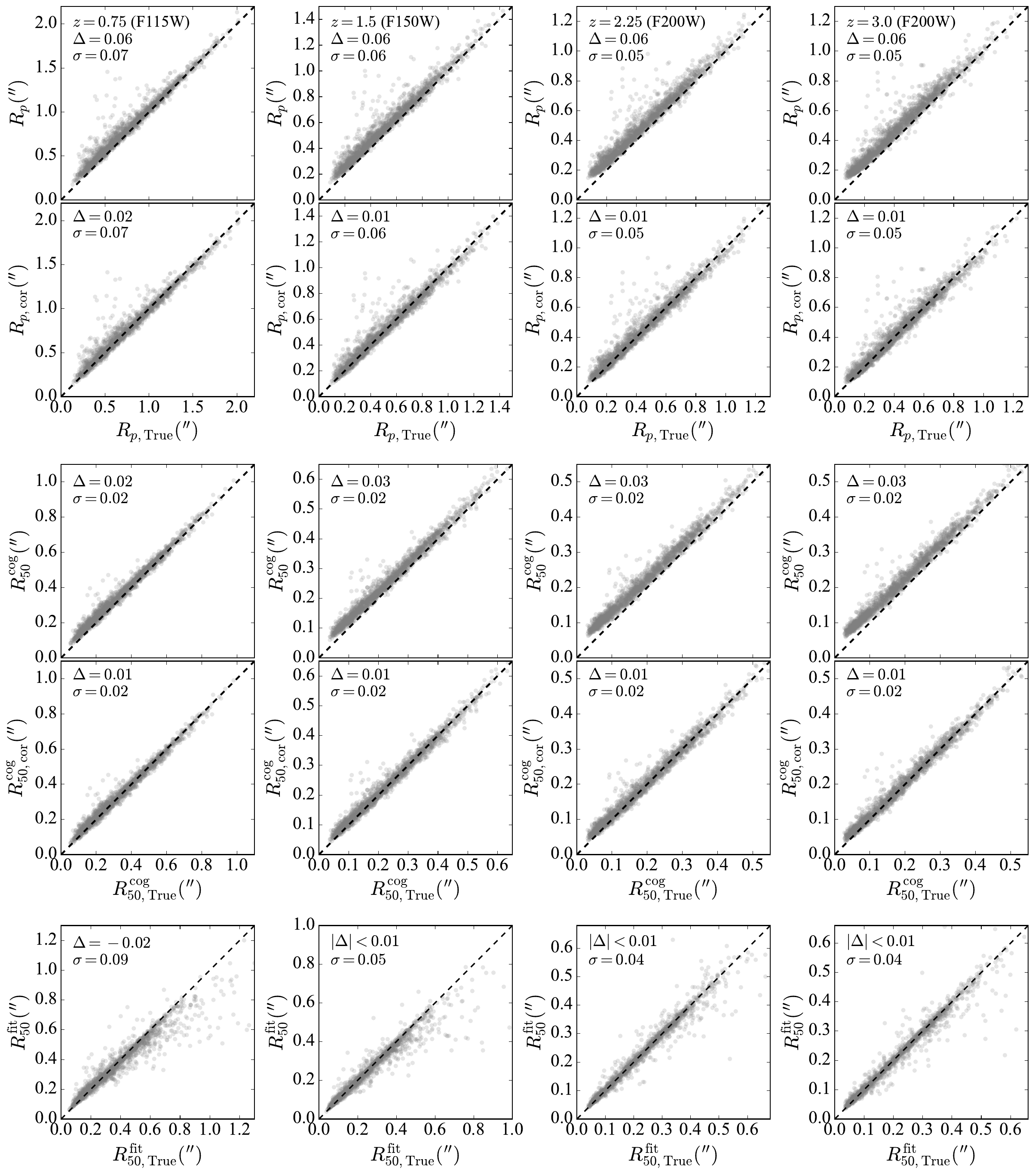}
        \caption{Comparison of galaxy sizes measured from simulated CEERS images, bias-corrected sizes, and intrinsic sizes. The top two rows present results for Petrosian radius ($R_p$). The middle two rows present results for half-light radius obtained through non-parametric method ($R^{\rm cog}_{50}$). The bottom row presents results for half-light radius obtained through Sérsic fitting ($R^{\rm fit}_{50}$), with no correction applied. The mean difference ($\Delta$) and scatter ($\sigma$) between the $x$-axis and $y$-axis values are indicated at the top of each panel. Columns one through four show results for $z=0.75$, 1.5, 2.25, and 3.0, respectively.
        }
        \label{Comp_s}
\end{figure*}
\begin{figure*}
        \centering
        \includegraphics[width=0.95\textwidth]{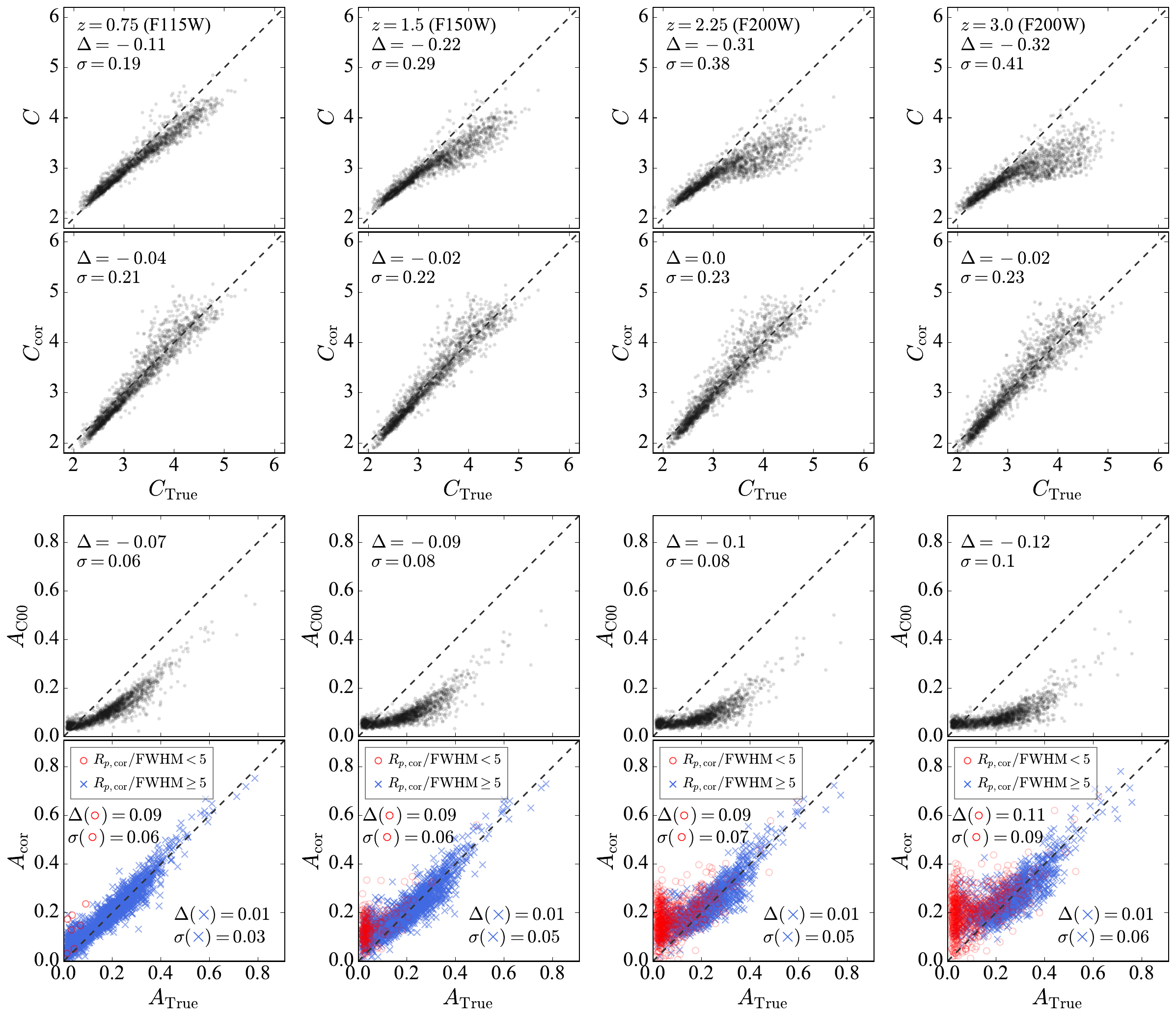}
        \caption{Evaluation of the effectiveness of the correction functions in reproducing the intrinsic concentration and asymmetry. 
        The top two rows present comparisons of the concentration measured from simulated CEERS images ($C$), bias-corrected values ($C_{\rm cor}$), and intrinsic values ($C_{\rm True}$). The bottom two rows illustrate comparisons of asymmetry measured from simulated CEERS images using the conventional method ($A_{\rm C00}$), bias-corrected values ($A_{\rm cor}$), and intrinsic values ($A_{\rm True}$). Columns one through four show results for $z=0.75$, 1.5, 2.25, and 3.0, respectively. The mean difference ($\Delta$) and scatter ($\sigma$) between the $y$-axis and $x$-axis values are indicated at the top of each panel. In the $A_{\rm True}$\textendash$A_{\rm cor}$ relations, data are divided into two groups: angularly small galaxies with $R_{p,\,\rm{True}}/{\rm FWHM}<5$ (red circles) and angularly large galaxies with $R_{p,\,\rm{cor}}/{\rm FWHM}\geq 5$ (blue crosses).
        }
        \label{Comp_CA}
\end{figure*}
\begin{figure*}
        \centering
        \includegraphics[width=0.95\textwidth]{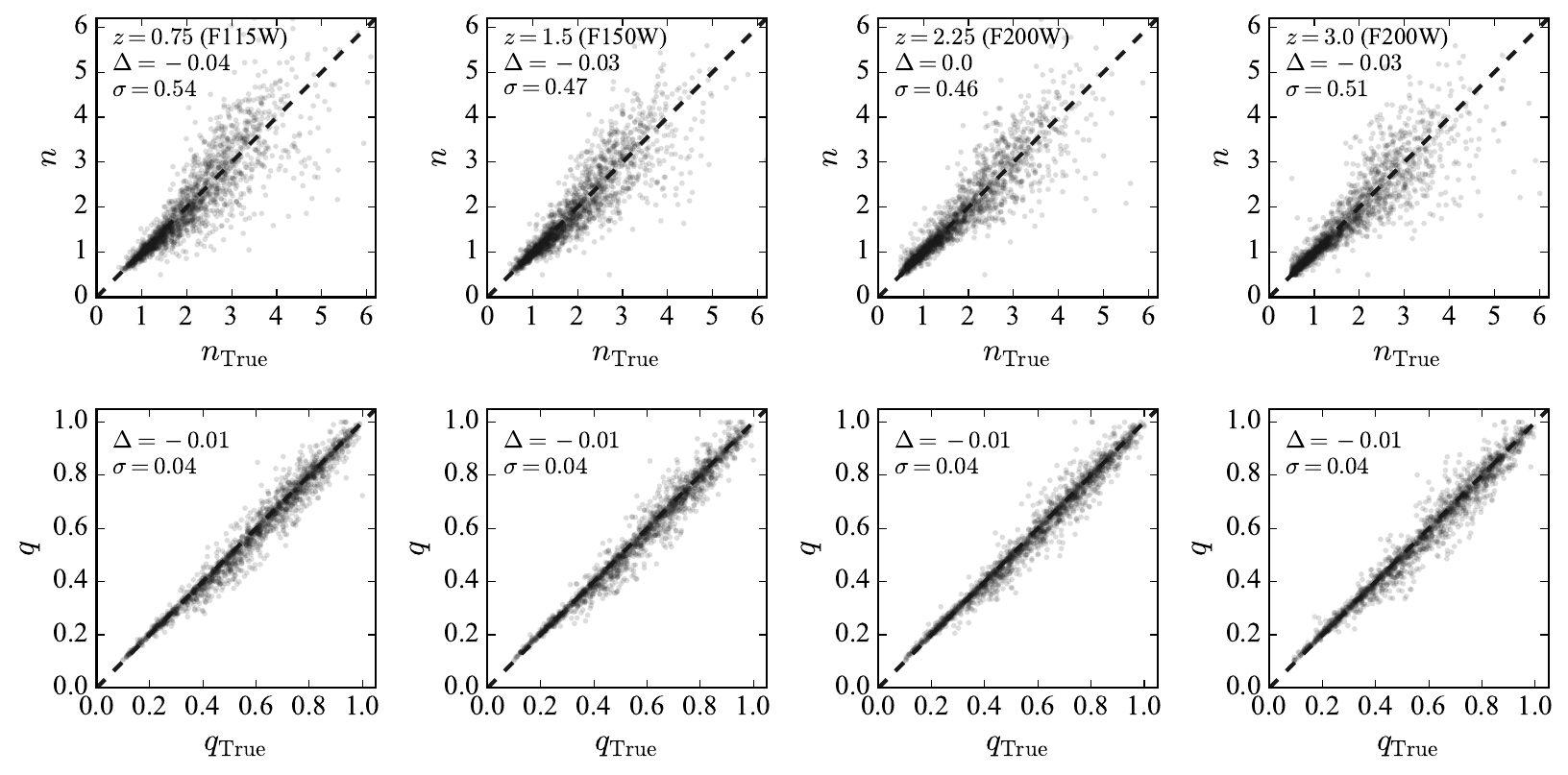}
        \caption{
        Evaluation of the effectiveness of the correction functions in reproducing the intrinsic Sérsic index and axis ratio.
        Top: Comparison between Sérsic index measured from the simulated CEERS images ($n$) and the intrinsic values ($n_{\rm True}$). Bottom: Comparison between axis ratio measured from the simulated CEERS images ($q$) and the intrinsic values ($q_{\rm True}$). The mean difference ($\Delta$) and scatter ($\sigma$) between the $y$-axis and $x$-axis values are indicated at the top of each panel.
        }
        \label{Comp_nq}
\end{figure*}

In this section, we describe how we apply the correction functions to the morphological quantities measured from the simulated CEERS images to understand the effectiveness of these functions. We start by measuring $R_p$, $R^{\rm cog}_{50}$, $R^{\rm fit}_{50}$, $C$, $A_{\rm C00}$, $n$, and $q$ on the simulated CEERS images. The corrections are summarized as follows. We correct size measurements using Eq.~(\ref{cor_Rp}) and (\ref{cor_R50}) to obtain $R_{p,\,{\rm cor}}$ and $R^{\rm cog}_{50,\,{\rm cor}}$, respectively.  The resolution level is calculated as $R_{p,\,\rm{cor}}/{\rm FWHM}$. We obtain the parameters $k$ and $b$ through Eq.~(\ref{corr_k}) and Eq.~(\ref{corr_b}), respectively, and then calculate the bias-corrected concentration ($C_{\rm cor}$) using the correction function Eq.~(\ref{cor_C}).  We calculate the noise-removed asymmetry using Eq.~(\ref{WZ}), with $f_1=2.25$ and $f_2=2.1$, obtain the parameters $T$, $D$, and $A_0$ through Eq.~(\ref{T}), (\ref{D}), and (\ref{A0}), respectively, and, finally, we calculate the bias-corrected asymmetry ($A_{\rm cor}$) using the correction function Eq.~(\ref{Acorf}). There is no correction for the parameters ($R^{\rm fit}_{50}$, $n$, and $q$) measured through Sérsic fitting using {\tt IMFIT}.  The statistical uncertainties are computed using the uncertainty function for each morphological measurement, but the derived uncertainties are not plotted in the figures of the following discussion for the sake of clarity.

In Fig.~\ref{Comp_s}, we plot $R_p$ and $R_{p,\,{\rm cor}}$ as a function of $R_{p,\,{\rm True}}$ in the top two rows, $R^{\rm cog}_{50}$ and $R^{\rm cog}_{50,\,{\rm cor}}$ as a function of $R^{\rm cog}_{50,\,{\rm True}}$ in the middle two rows, and $R^{\rm fit}_{50}$ as a function of $R^{\rm fit}_{50,\,{\rm True}}$ in the bottom row. Results for $z=0.75$, 1.5, 2.25, and 3.0 are presented in columns one through four, respectively, and the difference and scatter between $y$-axis and $x$-axis values are indicated in the top of each panel (Figs.~\ref{Comp_CA} and \ref{Comp_nq} follow the same strategy).  Values of $R_p$ and $R^{\rm cog}_{50}$ slightly overestimate their intrinsic values by 0.06\,arcsec and 0.03\,arcsec, respectively. After bias correction, these overestimations are reduced to 0.01\,arcsec, which is negligibly small. The correlations between $R^{\rm fit}_{50}$ and  $R^{\rm fit}_{50,\,{\rm True}}$ exhibit a very small average offset, but they display a slightly larger scatter compared with the non-parametric measures of galaxy sizes.

The top two rows of Fig.~\ref{Comp_CA} presents the plots of $C$ and $C_{\rm cor}$ as a function of $C_{\rm True}$. CEERS galaxies with higher $C_{\rm True}$ tend to have measured $C$ to be underestimated, and since the galaxies are angularly smaller at higher redshifts, the underestimation becomes more severe. Scatters are especially large at high redshifts.  As early-type galaxies are more centrally concentrated than late-type galaxies, the $C$ values of early-type galaxies will be more underestimated without correction.  After bias correction, $C_{\rm cor}$ correlates well with $C_{\rm True}$, exhibiting a small average offset of approximately $-0.02$ and with an average scatter of around 0.22.

The bottom two rows of Fig.~\ref{Comp_CA} presents the plots of $A_{\rm C00}$ and $A_{\rm cor}$ as a function of $A_{\rm True}$. Our results show that $A_{\rm C00}$ underestimate $A_{\rm True}$ on average, especially at high  $A_{\rm True}$ values or at high redshifts, due to the overestimation of noise contribution and the effects of PSF smoothing. Since late-type galaxies have higher intrinsic asymmetry than early-type galaxies \citep{Abraham1996}, the measured asymmetry for late-type galaxies (if not corrected) will be more severely underestimated compared to that of early-type galaxies. Our findings suggest that employing the conventional algorithm for calculating galaxy asymmetry, as proposed  in \cite{Conselice2000}, is unsuitable and inadequate to measure the intrinsic asymmetry of high-redshift galaxies observed in CEERS.  In a recent study, \cite{Kartaltepe2023} visually classified the morphological type of 850 galaxies at $z>3$ observed in JWST CEERS and used {\tt Statmorph} to perform morphological measurements. They used images from the filter corresponding to the rest-frame optical emission at the redshift of the galaxy. Their results reveal that the Concentration\textendash Asymmetry diagram does not clearly differentiate between morphological types, although peculiar galaxies exhibit slightly higher asymmetry, on average. Our findings provide a possible explanation to the results in \cite{Kartaltepe2023}: the CEERS PSF and noise introduce significant bias and uncertainty to the measurements of concentration and asymmetry, causing the resulting concentration\textendash asymmetry diagram to be degenerate across galaxy types.

The $A_{\rm True}$\textendash$A_{\rm cor}$ relations demonstrate that our improved method can accurately reproduce the intrinsic galaxy asymmetry when $z\lesssim1.5$. However, when $z \gtrsim 1.5$, the relations become increasingly dispersive with higher redshifts. We divide the simulated CEERS galaxies into two groups: angularly small galaxies with $R_{p,\,\rm{True}}/{\rm FWHM}<5$, marked with red circles, and angularly large galaxies with $R_{p,\,\rm{cor}}/{\rm FWHM}\geq 5$, marked with blue crosses. The results show that, for angularly large galaxies, $A_{\rm cor}$ reproduces $A_{\rm True}$ well with a mean difference of $0.01$ across all redshifts, and a scatter increasing from 0.03 at $z=0.75$ to 0.06 at $z=3.0$. In contrast, for angular small galaxies, the $A_{\rm cor}$ still overestimates $A_{\rm True}$ with a mean difference of approximately $0.1$ across all redshifts and with scatter increasing from 0.06 at $z=0.75$ to 0.09 at $z=3.0$. The latter is caused by the incomplete removal of noise correction for the most symmetric galaxies, even when using our improved noise correction (Fig.~\ref{CAsky}), and by the flatness of the correction function (Fig.~\ref{CApsf}). Therefore, our asymmetry correction function is only efficient for angularly large galaxies ($R_{p,\,\rm{cor}}/{\rm FWHM}\geq 5$) observed in JWST CEERS.

In Fig.~\ref{Comp_nq}, we plot $n$ against $n_{\rm True}$ in the top row and $q$ against $q_{\rm True}$ in the bottom row. The average difference between $n$ and $n_{\rm True}$ is small, approximately $-0.03$, with an average scatter of around 0.5. This suggest that the Sérsic fitting can extract the index without significant bias for galaxies observed in JWST CEERS, but the statistical uncertainty should be appropriately considered. The $q$\textendash$q_{\rm True}$ relations are quite tight with negligible offset and scatter, indicating that the Sérsic fitting can robustly extract the axis ratio.

\section{Summary and conclusions}\label{conclusions}

Early JWST studies have shown that galaxies with established disk and spheroidal morphologies span the full redshift range, revealing that the Hubble sequence was already in place at the early universe \citep[e.g.,][]{Ferreira2022a, Ferreira2022b, Kartaltepe2023, Nelson2022, Robertson2023}. However, potential biases and uncertainties in characterizing the morphologies of these high-redshift galaxies may significantly impact the results. To address this issue, we defined a sample of nearby galaxies based on the DESI survey, conststing of 1816 galaxies with stellar masses ranging from $10^{9.75}\,M_{\odot}$ to $10^{11.25}\,M_{\odot}$. High-quality DESI images of these nearby galaxies allow for an accurate determination of the galaxy morphology. We removed the contamination from foreground stars or background galaxies in the DESI images and used the resulting cleaned images of the {\it g}, {\it r}, and {\it z} bands as input to compute artificial images of galaxies located at $0.75\leq z\leq 3$ and observed at rest-frame optical wavelengths in CEERS. We used the F115W filter for $z=0.75$ and 1, the F150W filter for $z=1.25$, 1.5, and 1.75, and the F200W filter for $z=2.0$ to 3.0. For the artificially redshifting process, we take into account angular size changes due to distance and intrinsic evolution, flux changes due to cosmological dimming and intrinsic evolution, spectral change, and changes in resolution and noise level. The rest-frame flux is obtained by pixel-by-pixel {\it K} correction. The simulated images are binned onto the pixel scale of 0.03\,arcsec$/$pixel and corrected to the JWST PSF. A patch of real CEERS background and poisson noise from galaxy light are then added.

We focus on quantifying the biases and uncertainties for six widely used morphological quantities: Petrosian radius ($R_p$), half-light radius ($R_{50}$), asymmetry ($A$), concentration ($C$), axis ratio ($q$), and Sérsic index ($n$). The resolution of CEERS images emerges as the primary factor influencing these measurements, while the typical CEERS noise predominantly impacts the computation of $A$. We improve the method for removing the contribution of noise from the computation of $A$. To correct for the resolution effects, we reduce the DESI image resolution, conduct each measurement, and compare these re-measured values with their intrinsic values to understand resolution effects and to derive formulae as a function of resolution level, defined as $R_p/{\rm FWHM}$, for correcting the biases and uncertainties. Finally, we apply the correction functions to our artificially redshifted CEERS images to validate the methods. Our main results are as follows.

\begin{enumerate}
  \item $R_p$ and $R_{50}$, measured using non-parametric approaches, are slightly overestimated due to PSF smoothing, and this overestimation does not significantly depends on the resolution level. We derived Eq.~(\ref{cor_Rp}) and  Eq.~(\ref{cor_R50}) to correct these biases. In comparison, $R_{50}$ measured using 2D image fitting proves to be unbiased.  The functions of statistical uncertainties for these three parameters are provided. 
  \item $C$ is underestimated due to PSF smoothing, with the effect being more pronounced for higher $C$ values or lower resolutions. The bias can be corrected using correction function of Eq.~(\ref{cor_C}) and the  statistical uncertainty is given by Eq.~(\ref{sig_C}). 
  \item By incorporating a more accurate noise effect removal procedure, we improved the computation of $A$ over existing methods, which can often overestimate, underestimate, or lead to significant scatter of noise contributions.  We show that $A$ of the most intrinsically symmetric galaxies are overestimated due to the PSF asymmetry. In contrast, $A$ of intrinsically asymmetric galaxies are underestimated owing to smoothing, particularly for large $A$ values and at lower resolutions. 
 For angularly large CEERS galaxies where $R_p/{\rm FWHM}\geq 5$, the biases can be robustly corrected using Eq.~(\ref{Acorf}). However, for smaller galaxies, these biases cannot be completely removed. When studying asymmetry, the statistical uncertainty given by Eq.~(\ref{sig_A}) should be taken into account. 
  \item The measurements of $n$ and $q$ through 2D image fitting have negligible biases. The statistical uncertainty for axis ratio is negligible, whereas the uncertainty for the Sérsic index is more significant and should be properly considered.
  \item Although our primary focus is on studying the optical morphology of simulated galaxies at $z\leq3$ observed with F115W, F150W, and F200W filters, we also provide the correction for F277W, F356W, and F444W filters, which may be useful for studying optical morphology of galaxies at higher redshifts. The parameters of the correction functions vary across different filters and are provided in Tables~\ref{best-fit} and \ref{best-fit2}.
\end{enumerate}

These tests establish a solid foundation for future quantitative statistical studies aimed at comprehending the cosmological evolution of galaxy morphology. The dataset of artificially redshifted images also holds significant value for additional studies, such as those focused on spiral arms and bars.

\begin{acknowledgements}
We are grateful to the anonymous referee for their invaluable feedback and insightful comments that greatly improved the quality of this paper. 
SYY acknowledges the support by the Alexander von Humboldt Foundation. 
SYY thank Luis C. Ho for inspiring him to pursue research on high-redshift galaxies. 
We thank the fruitful discussion with John Moustakas. We thank Song Huang for creating the Slack workspace that enabled SYY to find collaboration with CC and FS.\\
This research made use of Photutils, an Astropy package for
detection and photometry of astronomical sources \citep{Bradley2022}.
We acknowledge the usage of the HyperLeda database (http://leda.univ-lyon1.fr).\\

This work is based on observations taken by the 3D-HST Treasury Program (GO 12177 and 12328) with the NASA/ESA HST, which is operated by the Association of Universities for Research in Astronomy, Inc., under NASA contract NAS5-26555.\\

The Legacy Surveys consist of three individual and complementary projects: the Dark Energy Camera Legacy Survey (DECaLS; Proposal ID \#2014B-0404; PIs: David Schlegel and Arjun Dey), the Beijing-Arizona Sky Survey (BASS; NOAO Prop. ID \#2015A-0801; PIs: Zhou Xu and Xiaohui Fan), and the Mayall z-band Legacy Survey (MzLS; Prop. ID \#2016A-0453; PI: Arjun Dey). DECaLS, BASS and MzLS together include data obtained, respectively, at the Blanco telescope, Cerro Tololo Inter-American Observatory, NSF’s NOIRLab; the Bok telescope, Steward Observatory, University of Arizona; and the Mayall telescope, Kitt Peak National Observatory, NOIRLab. Pipeline processing and analyses of the data were supported by NOIRLab and the Lawrence Berkeley National Laboratory (LBNL). The Legacy Surveys project is honored to be permitted to conduct astronomical research on Iolkam Du’ag (Kitt Peak), a mountain with particular significance to the Tohono O’odham Nation.

NOIRLab is operated by the Association of Universities for Research in Astronomy (AURA) under a cooperative agreement with the National Science Foundation. LBNL is managed by the Regents of the University of California under contract to the U.S. Department of Energy.

This project used data obtained with the Dark Energy Camera (DECam), which was constructed by the Dark Energy Survey (DES) collaboration. Funding for the DES Projects has been provided by the U.S. Department of Energy, the U.S. National Science Foundation, the Ministry of Science and Education of Spain, the Science and Technology Facilities Council of the United Kingdom, the Higher Education Funding Council for England, the National Center for Supercomputing Applications at the University of Illinois at Urbana-Champaign, the Kavli Institute of Cosmological Physics at the University of Chicago, Center for Cosmology and Astro-Particle Physics at the Ohio State University, the Mitchell Institute for Fundamental Physics and Astronomy at Texas A\&M University, Financiadora de Estudos e Projetos, Fundacao Carlos Chagas Filho de Amparo, Financiadora de Estudos e Projetos, Fundacao Carlos Chagas Filho de Amparo a Pesquisa do Estado do Rio de Janeiro, Conselho Nacional de Desenvolvimento Cientifico e Tecnologico and the Ministerio da Ciencia, Tecnologia e Inovacao, the Deutsche Forschungsgemeinschaft and the Collaborating Institutions in the Dark Energy Survey. The Collaborating Institutions are Argonne National Laboratory, the University of California at Santa Cruz, the University of Cambridge, Centro de Investigaciones Energeticas, Medioambientales y Tecnologicas-Madrid, the University of Chicago, University College London, the DES-Brazil Consortium, the University of Edinburgh, the Eidgenossische Technische Hochschule (ETH) Zurich, Fermi National Accelerator Laboratory, the University of Illinois at Urbana-Champaign, the Institut de Ciencies de l’Espai (IEEC/CSIC), the Institut de Fisica d’Altes Energies, Lawrence Berkeley National Laboratory, the Ludwig Maximilians Universitat Munchen and the associated Excellence Cluster Universe, the University of Michigan, NSF’s NOIRLab, the University of Nottingham, the Ohio State University, the University of Pennsylvania, the University of Portsmouth, SLAC National Accelerator Laboratory, Stanford University, the University of Sussex, and Texas A\&M University.

BASS is a key project of the Telescope Access Program (TAP), which has been funded by the National Astronomical Observatories of China, the Chinese Academy of Sciences (the Strategic Priority Research Program “The Emergence of Cosmological Structures” Grant \# XDB09000000), and the Special Fund for Astronomy from the Ministry of Finance. The BASS is also supported by the External Cooperation Program of Chinese Academy of Sciences (Grant \# 114A11KYSB20160057), and Chinese National Natural Science Foundation (Grant \# 12120101003, \# 11433005).

The Legacy Survey team makes use of data products from the Near-Earth Object Wide-field Infrared Survey Explorer (NEOWISE), which is a project of the Jet Propulsion Laboratory/California Institute of Technology. NEOWISE is funded by the National Aeronautics and Space Administration.

The Legacy Surveys imaging of the DESI footprint is supported by the Director, Office of Science, Office of High Energy Physics of the U.S. Department of Energy under Contract No. DE-AC02-05CH1123, by the National Energy Research Scientific Computing Center, a DOE Office of Science User Facility under the same contract; and by the U.S. National Science Foundation, Division of Astronomical Sciences under Contract No. AST-0950945 to NOAO. \\

The Siena Galaxy Atlas was made possible by funding support from the U.S. Department of Energy, Office of Science, Office of High Energy Physics under Award Number DE-SC0020086 and from the National Science Foundation under grant AST-1616414.

\end{acknowledgements}

\end{document}